\documentclass[11pt]{article}
\usepackage{mathrsfs}
\usepackage{amsfonts}
\usepackage{amsmath,amsthm,hyperref}
\usepackage{amssymb}
\usepackage{hyperref,color}
\usepackage{exscale,relsize}
\usepackage[all]{xy}
\usepackage{exscale,relsize,graphicx,caption}

\makeatletter
\def\equfill@{\arrowfill@\Relbar\Relbar\Relbar}
\newcommand{\equfill}[2][]{\ext@arrow 0395\equfill@{#1}{#2}}
\makeatother
\makeatletter
\def\Eqlfill@{\arrowfill@\Relbar\Relbar\Relbar}
\newcommand{\extendEql}[1][]{\ext@arrow 0359\Eqlfill@{#1}}
\makeatother

\numberwithin{equation}{section}
\begin{document}

\author{ Peng Zhao   \and  Engui Fan  \footnote{Corresponding
author and  e-mail address:
      faneg@fudan.edu.cn} \and }
\date{   \small{ School of Mathematical Sciences, Institute of Mathematics \\
 and Key Laboratory of Mathematics for Nonlinear Science, \\ Fudan
University, Shanghai 200433, P.R. China} }
\title{\bf \Large{ Algebro-Geometric Solutions for Kadomtsev-Petviashvili Hierarchy} }
\maketitle
\begin{abstract}
 Based on the idea of symmetric constraint,
we apply the Gesztesy-Holden's   method
to derive explicit representations of the Baker-Ahkiezer function $\psi_1$ of the KP hierarchy, from which  we  provide theta function representations of
algebro-geometric solutions for the whole
Kadomtsev-Petviashvili (KP) hierarchy. This provides a approach to   obtain   some special subclasses of algebro-geometric solutions for the KP
 hierarchy and other high dimensional hierarchy of equations.

\end{abstract}

\section{Introduction}
In the past few decades there have been many remarkable developments in the theory of
integrable nonlinear partial differential equations \cite{as1981},~\cite{Belokolos1994},~\cite{Dickey1991},~\cite{zms1980}. One of
the most widely studied integrable
equation in 2+1 dimensions is
the Kadomtsev-Petwiashvili (KP) equation \cite{dkjm1983},~\cite{JM1983},
\begin{equation}\label{1.1kp}
  u_t=\frac{1}{4}u_{xxx}+3uu_x+\frac{3}{4}\partial^{-1}_xu_{yy},
\end{equation}
which also has been generalized in   various forms \cite{mtj1992},~\cite{ya},~\cite{zms1980}.
There are also
different approaches to the description of the algebraical and geometrical aspects of the
KP equation.  One is the theory of bilocal recursion operators \cite{asfokas},~\cite{Konopelchenko1},~\cite{santini}, where the KP
appears as a member of a hierarchy of commuting flows. Another is the Sato's theory \cite{dmmt1981},~\cite{dkjm1983},~\cite{JM1983},~\cite{sato}, which is based on the treatment of partial differential KP equations as dynamical systems on the infinite-dimensional algebra of pseudo-differential operators \cite{a1}. By introducing an infinite set of ¡®time¡¯ variables one can also treat the integrable
equations as flows on infinite-dimensional Grassmannian manifolds \cite{JM1983}. Moreover, this
theory reveals deep interrelations between the Hamiltonian structures of the KP hierarchy
and two-dimensional conformal field theory as well as W$_{1+\infty}$ algebras~\cite{Yamagishi}.
In the scheme of the Sato's theory,
the KP equation is considered as the simplest member of a hierarchy of
equations which can be brought to bilinear form and solved by the $\tau$-funtion approach.


The problem of constructing the algebro-geometric solutions  is one of the most challenging
problems of the theory of integrable systems, and many mathematicians and physicists
spent much efforts to obtain the algebro-geometric solutions  for almost all equations that are known to be integrable.
This kinds of solutions were originally studied on the KdV equation based on
the inverse spectral theory and algebro-geometric method developed by pioneers such as Novikov~\cite{DubNov}, Dubrovin \emph{et al.}~\cite{dmn1976},~\cite{Novikov1976m}, Its \& Matveev~~\cite{itsma1975},~\cite{Its1975}, Lax~\cite{lax1975},
and McKean \& van Moerbeke~\cite{mckean1975} for 1 + 1 systems, and
extended by Krichever in 1976 for 2 + 1 systems like KP~\cite{Krichever1976},~\cite{Krichever1977a} in the late 1970s.
Later this theory has
been developed to the whole hierarchies of nonlinear integrable equations by Gesztesy,
Holden \emph{et al.} using polynomial recursion method \cite{gesztesy1999},~\cite{gesztesy1998},~\cite{gesztesy2003a},~\cite{Gesztesyhoden2003}.
 Another breakthrough in this area was made by Mumford~\cite{Mumford} in early 1980s, who observed  that integrable equations like KdV, KP or sine-Gordon, are hidden in Fay's trisecant formula. Mumford's approach are based on degenerated versions
of Fay's identity, which reveals the relations between algebro-geometric solutions of integrable equations and a purely algebro-geometric identity  \cite{ckalla2013}, \cite{clein2002}, \cite{clein2005}. Owing to their work, the theory of theta functions and Abelian varieties developed in
the domain of complex analysis and
algebraic geometry can be directly linked with the theory of integrable systems.
A detailed introduction to this aspect can be found in the survey article \cite{Matveev2008}.


Before turning to the main text, it seems appropriate to review some related literature
about KP equation in detail
as usual.
An important development of the algebro-geometric method
was the passage from 1+1 systems to the integration of 2+1 KP-like systems, realized by Krichever
in 1976 \cite{Krichever1976}, who constructed algebro-geometric solutions of
the KP eqation on a
basis of a purely algebraic formulation of algebro-geometric approach.
A complete description of smooth, real, algebro-geometric solutions of the
KP equation was obtained in \cite{Dubrovin1989m} by separate nontrival work.
The algebro-geometric spectral theory of the stationary periodic
2D Schr\"{o}dinger operator was developed in the works \cite{Novikov1976m}, \cite{Novikov1976n}.
In ref. \cite{miller}, Miller found 2+1-dimensional dynamics of the equations of the KP hierarchy can be approximated by the simpler 1+1-dimensional dynamics of the equations of a sequence of larger and larger vector nonlinear Schr\"{o}dinger hierarchies, which could be used to develop powerful numerical methods for
solving the KP equations. A explicit theta function solution of the KP
equation was derived with the help of separation technique by Cao \emph{et al.}~\cite{caogeng}.
Matveev \& Salle found
larger families of solutions to the KP equation, also expressed by means of the
Riemann theta functions, by applying the dressing formulae \cite{Matveev1979mm}, \cite{Matveev1979mn}.

The purpose of this paper is to construct the algebro-geometric solutions for the whole KP hierarchy
by extending the Gesztesy-Holden's  method of \cite{Gesztesyhoden2003} to 2+1 dimensional case based on the idea of
symmetry constraint. The KP hierarchy can be consistently constrained in many different ways to yield hierarchies of equations in 1 + 1 independent variables \cite{chengyi},~\cite{cheng1992},~\cite{Fokas1986},~\cite{Konopelchenko1992a},~\cite{miller},\cite{Oevel1993},~\cite{Sidorenkot1991},~ \cite{santini}, ~\cite{xubing1993}. The idea of studying these constraints comes from the reduction of 1+1-dimensional integrable soliton
equations to finite-dimensional integrable equations \cite{Antonowitcz1991},~\cite{Flaschka1983},~\cite{Gardner1974}.
A well-known example is the restriction of the KdV
flow to the pure multisoliton submanifold \cite{Gardner1974}, where we impose the constraint
$u=\sum_{i=1}^N c_i\Psi_i^2$ on the KdV
potential $u$ and eigenfunctions $\Psi_i.$
This leads to the finite-dimensional integrable system
$$\Psi_{i,xx}=\lambda_i^2\Psi_i+\sum_{k=1}^Nc_k\Psi_k^2\Psi_i, i=1,\ldots,N.$$
 For the KP equations,
there are two kinds of constraints.
The one is
the so-called $k$-reductions for which the operator $L^k$ is forced
to become a purely differential operator leads to hierarchies
of $1+1$-dimensional equations \cite{dkjm1983}, \cite{JM1983}.
These reductions
are associated with constraints imposed separately
on the potentials $u$, and eigenfunctions $\psi$ and $\psi^*$. Meanwhile,
it has been shown that the KP hierarchy also
admits another type of constraints relating
the potenials $u$, with the eigenfunctions $\psi$ and $\psi^*$. One of them is
the symmetry constraints for 2+1 dimensional soliton equations, which have been discussed
the first time in the papers \cite{chengyi}, \cite{Konopelchenko1991}.
In ref. \cite{Konopelchenko1991}, Konopelchenko \emph{et al.} proved that the auxiliary linear problems
\begin{equation}\label{st1.2}
\begin{split}
&\psi_{t_2}=\psi_{xx}+2u \psi,    \\                                                                                    &\psi_{t_3}=\psi_{xxx}+ 3u\psi_x+\frac{3}{2}u_x\psi+\frac{3}{2}\partial^{-1}_xu_{t_2}\psi,
\end{split}
\end{equation}
and its adjoint
\begin{equation}\label{st1.3}
  \begin{split}
 &  \psi_{t_2}^*=-\psi_{xx}^*-2u \psi^*,\\
 & \psi_{t_3}^*=\psi_{xxx}^*+ 3u\psi_x^*+\frac{3}{2}u_x\psi^*-\frac{3}{2}\partial^{-1}_xu_{t_2}\psi^*
  \end{split}
\end{equation}
which arise from the third flow and second flow of KP hierarchy
are constrained to
the first two nontrivial flows of the AKNS hierarchy
by identifying the potential $u$
of the KP equation to $\psi\psi^*$.
Further, it has been shown that imposing the symmetry constraint
\begin{equation*}
  u_x=(\psi\psi^*)_x
\end{equation*}
on the auxiliary linear problems of the KP hierarchy
\begin{align}
\psi_{t_m}=L_+^m\psi,~~
\psi_{t_m}^*=-(L_+^m)^*\psi,
\end{align}
leads to standard AKNS hierarchy \cite{Konopelchenkot1991}. Here
$L$ is the usual
first-order formal pseudo-differential Lax operator defined in (\ref{sato1})
and $(L_+^m)^*$ is the formal adjoint to the operator $L_+^m,$ i.e. if $L_+^m=\sum v_j\partial_x^j$,
then $(L_+^m)^*=\sum (-\partial_x)^jv_j.$ Further motivation for the new constraints lies in the methods of solving integrable equations
in 1+1- or 2+l-dimensions by the nonlinearization of linear problems \cite{cao1990},~\cite{cheng1991},~\cite{zeng1990}.
Thus, imposing some constraint on a 2+1
  dimensional equation
  is a possible way to obtain a submanifold of solutions of the KP equations  by solving two equations in 1+1 dimensions.

Since
any non-singular algebraic
curve can be used as spectral curve of KP equation, it is essential to
consider some constraints about KP hierarchy if we really want
to get a explicit form of Baker-Akhiezer function from this point of view.
However, this way of consideration produces mixed results. On the one hand,
the study of KP equations is restricted to only certain specific types of spectral curves, which
narrows the classes of possible solutions. On the other hand, it makes
problems more concrete, which can increase the expressions of solutions
corresponding to this specific curve
and allows us to have a unified way
to study solutions of the whole KP hierarchy. Thus,
considering the constraints of KP equations enables us to get
larger families of algebro-geometric solutions
in this sense, which is also our start point.

This paper is organized as follows.
In section 2, we formulate some fundamental knowledge about Sato's KP hierarchy,
AKNS hierarchy and the relations between them in literature. Then a basic initial problem
is introduced as solutions of a linear system which is defined by classical squared basis functions.
In section 3, we shall introduce hyperellipitc curves associated with KP hierarchy and
study
the dynamics of auxiliary spectral points $\{\mu_j\}_{j=1}^n, \{\nu_j\}_{j=1}^n$ with
respect to $x,y,t_{r+1}$ and corresponding trace formula. In section 4, we present explicit
representation for
Baker-Akhiezer function $\psi_1, \psi_2$ and consider their analytic properties. In section 5, it will be shown the function $\psi_1$
derived in section 4 is not only connected to the Baker-Akhiezer function of the KP equation,
but a quantity to define new 2+1 dimension systems possessing close relations with the KP equation.
Moreover, we shall derive
theta function representation for $\psi_1,\psi_2, q,p$,
and algebro-geometric solutions of the whole KP hierarchy.

The whole approach discussed in the present paper is a general one and gives
similar results for other 2+1 dimensional and higher dimensional soliton equations, for instance, for the
modified KP hierarchy and
Davey-Stewartson equation, etc.

\section{Sato KP hierarchy, AKNS hierarchy and basic initial value problem}

In the Sato approach~\cite{dkjm1983}, the KP hierarchy is described by the
isospectral deformations of the eigenvalue problem
\begin{equation}\label{st2.1}
\begin{split}
 & L\psi=\lambda \psi,~~ \lambda\in\mathbb{C},
  \end{split}
\end{equation}
where the pseudodifferential operator $L$ is given by
\begin{equation}\label{sato1}
L= \partial+ \sum_{j=1}^{\infty} u_{j+1}\partial^{-j},~~ \partial=\partial_x,
\end{equation}
and $u_j$ are functions in
infinitely many variables $(t_1,t_2,\ldots)$ with $t_1=x, t_2=y.$
We denote by $B_m$ the differential part of $L^m:$
\begin{align}\label{sato2}
   B_m= (L^m)_+=\sum_{j=0}^m b_{m,j}\partial^j.
\end{align}
The coefficients $b_{m,j}$ in (\ref{sato2}) can be uniquely determined
by the coordinates $u_j$, and their $x$ derivatives.
Explicitly,
\begin{align}
  B_1=&~\partial, \nonumber\\
  B_2=&~\partial^2+2u_2,\nonumber\\
  B_3=&~\partial^3+3u_2\partial+3 u_3+3u_{2,x},\nonumber\\
  B_4=&~\partial^4+4u_2\partial^2+(4 u_3+6u_{2,x})\partial+4u_4+6u_{3,x}\nonumber\\
   &+4u_{2,xx}+6u_2^2,~~\textrm{etc.} \nonumber
\end{align}
From the compatibility conditions of (\ref{st2.1}) and
\begin{equation}
  \phi_{t_m}=B_m\phi,
\end{equation}
we have
\begin{align}\label{st2.5}
  L_{t_m}=[B_m,L],
\end{align}
or equivalently,
\begin{equation}\label{st2.6}
  (B_{m})_{t_n}-(B_{n})_{t_m}=[B_n,B_m].
\end{equation}
The KP hierarchy is obtained from (\ref{st2.5}) or (\ref{st2.6}) for
infinite coordinates $\{u_j\}_{j=2}^\infty.$ For example,
from (\ref{st2.5}), we derive
\begin{align}
u_{2,t_2}=&~2u_{3,x}+u_{2,xx},\label{st2.7}\\
u_{3,t_2}=&~2u_{4,x}+u_{3,xx}+2u_2u_{2,x},\label{st2.8}\\
u_{4,t_2}=&~2u_{5,x}+u_{4,xx}+4u_{2,x}u_3-2u_2u_{2,xx},\label{st2.9}\\
&\ldots\ldots\nonumber\\
u_{2,t_3}=&~3u_{4,x}+3u_{3,xx}+u_{2,xxx}+6u_2u_{2,x},\label{st2.10}\\
u_{3,t_3}=&~3u_{5,x}+3u_{4,xx}+u_{3,xxx}+6(u_2u_{3})_x,\\
&\ldots\ldots\nonumber\\
u_{2,t_4}=&~4u_{5,x}+6u_{4,xx}+4u_{3,xxx}+u_{2,xxxx}+12(u_2u_3)_x\nonumber\\
&~+6(u_2u_{2,x})_x, \label{st2.12}\\
&\ldots\ldots\nonumber
\end{align}
Eliminating $u_3$, $u_4$, $u_5$ from (\ref{st2.7}), (\ref{st2.8}), (\ref{st2.9}) and
taking into account
(\ref{st2.10}), (\ref{st2.12}), one obtains
   \begin{align}
      &u_{t_3}=\frac{1}{4}u_{xxx}+3uu_x+\frac{3}{4}\partial^{-1}u_{yy}\label{stkp}\\
      &u_{t_4}=\frac{1}{2}u_{xxy}+4uu_y+2u_x\partial^{-1}u_y+\frac{1}{2}\partial^{-2}u_{yyy}.\label{stkp2}
\end{align}
where we denote by $u=u_2.$ Equation (\ref{stkp}) is just
the KP equation and equation (\ref{stkp2}) is the first higher order flows of the KP hierarchy.
Similarly, the KP and higher-order KP equation can also be derived from (\ref{st2.6}) with
$n=2,m=3$ and $n=2, m>3$, respectively.

Now we recall some basic results about AKNS hierarchy. The ref. \cite{Gesztesyhoden2003} provides two complexified
versions of AKNS hierarchy, zero-curvature and matrix differential operator formalisms. In the following, we shall construct the standard AKNS hierarchy in a similar manner.
Generally, the AKNS hierarchy is introduced by developing its
zero-curvature formalism. To this end
one defines the sequences of differential polynomials $\{f_\ell\}_{\ell\in\mathbb{N}_0}, \{g_\ell\}_{\ell\in\mathbb{N}_0},$ and $\{h_\ell\}_{\ell\in\mathbb{N}_0},$ recursively by
 \begin{align}
  f_0=&~-q,~g_0=-1/2,~h_0=p, \label{st2.13}\\
  f_{\ell+1}=&-f_{\ell,x}+2qg_{\ell+1}, ~\ell\in\mathbb{N}_0,\\
g_{\ell+1,x}=&~pf_{\ell}+qh_{\ell},~\ell\in\mathbb{N}_0, \\
h_{\ell+1}=&~h_{\ell,x}-2pg_{\ell+1},~\ell\in\mathbb{N}_0.\label{st2.16}
 \end{align}
Here we emphasize that $q,p$ should be regarded as functions of infinitely many variables $(t_{1}, t_{2},t_{3},\ldots).$
Explicitly, one computes
\begin{align*}
  &f_0=-q, \\
  &f_1=q_x+c_1(-q),\\
  &f_2=-q_{xx}+2pq^2+c_1 q_x+c_2(-q), \\
  &\ldots\ldots,\\
  &g_0=-1/2,\\
  &g_1=c_1(-1/2),\\
  &g_2=-pq+c_2(-1/2),\\
  &g_3=pq_x-p_xq+c_1(-pq)+c_3(-1/2),\\
  &\ldots\ldots,\\
  &h_0=p,\\
  &h_1=p_x+c_1 p,\\
  &h_2=p_{xx}-2p^2q+c_1 p_x+c_2 p,~~\textrm{etc.,}
\end{align*}
where $\{c_{\ell}\}_{\ell=0}^\infty\subset\mathbb{C}$ are integration constants.
By introducing the homogeneous coefficients $\hat{f}_\ell,\hat{g}_{\ell},\hat{h}_\ell$ by
vanishing of the integration constants $c_k, k=1,\ldots,\ell,$
\begin{align}
 &\hat{f}_0=-q, ~~\hat{f}_\ell=f_\ell|_{c_k=0,k=1,\ldots,\ell}\\
 &\hat{g}_0=-1/2,~~\hat{g}_\ell=g_\ell|_{c_k=0,k=1,\ldots,\ell}\\
 &\hat{h}_0=p,~~\hat{h}_\ell=h_\ell|_{c_k=0,k=1,\ldots,\ell}
\end{align}
we have
\begin{equation}
  f_\ell=\sum_{k=0}^\ell c_{\ell-k}\hat{f}_k,~~ g_\ell=\sum_{k=0}^\ell c_{\ell-k}\hat{g}_k,~~ h_\ell=\sum_{k=0}^\ell c_{\ell-k}\hat{h}_k,~~c_0=1.
\end{equation}
Then one
introduces
 \begin{align}
   U(z)=&~
    \left(
      \begin{array}{cc}
        -\frac{z}{2} & q \\
        p & \frac{z}{2} \\
      \end{array}
    \right),~~z\in\mathbb{C},\label{2.1}~~\\
    V_{n+1}(z)=&~\left(
                    \begin{array}{cc}
                      -G_{n+1}(z) & F_n(z) \\
                      -H_n(z) & G_{n+1}(z) \\
                    \end{array}
                  \right),~~n\in\mathbb{N}_0,\label{2.2}
 \end{align}
where
 $G_{n+1},F_n,H_n$ are polynomials with respect to $z:$
 \begin{align}
   G_{n+1}(z)=&~\sum_{\ell=0}^{n+1}g_{n+1-\ell}z^{\ell},\label{st2.23}\\
   F_n(z)=&~\sum_{\ell=0}^nf_{n-\ell}z^{\ell},\label{st2.24}\\
   H_n(z)=&~\sum_{\ell=0}^nh_{n-\ell}z^{\ell}.\label{st2.25}
 \end{align}
 and corresponding homogeneous polynomials
   are defined by
 \begin{align}
  &\widehat{F}_0(z)=F_0(z)=-q,~~\nonumber\\
  &\widehat{F}_\ell(z)=F_\ell(z)|_{c_k=0,k=1,\ldots,\ell}=\sum_{k=1}^\ell\hat{f}_{\ell-k}z^k,~~\ell\in\mathbb{N},\\
  &\widehat{G}_0(z)=G_0(z)=-1/2,~~\nonumber\\
  &\widehat{G}_{\ell+1}(z)=G_{\ell+1}(z)|_{c_k=0,k=1,\ldots,\ell+1}=\sum_{k=1}^{\ell+1}\hat{f}_{\ell+1-k}z^k,~~\ell\in\mathbb{N},\\
  &\widehat{H}_0(z)=H_0(z)=p,~~\nonumber\\
  &\widehat{H}_\ell(z)=H_\ell(z)|_{c_k=0,k=1,\ldots,\ell}=\sum_{k=1}^\ell\hat{f}_{\ell-k}z^k,~~\ell\in\mathbb{N}.
 \end{align}
 For fixed $n\in\mathbb{N}_0$, the stationary and time-dependent AKNS hierarchy are defined by demanding zero curvature equation
\begin{equation}\label{2.3}
V_{n+1,x}(z)=[U(z),V_{n+1}(z)]
\end{equation}
and
\begin{equation}\label{2.4}
U_{t_{n+1}}-V_{n+1,x}(z)=[U(z),V_{n+1}(z)],
\end{equation}
respectively.  Equation (\ref{2.3}) is equivalent to
\begin{align}
 0=&~-V_{n+1,x}+[U(z),V_{n+1}(z)]\nonumber\\
 =&~\left(
      \begin{array}{cc}
        G_{n+1,x}-pF_n-qH_n & -F_{n,x}-zF_n+2qG_{n+1} \\
        H_{n,x}-zH_n-2pG_{n+1}& -G_{n+1,x}+pF_n+qH_n \\
      \end{array}
    \right)\nonumber\\
    =&\left(
       \begin{array}{cc}
          0 & -f_{n+1} \\
          -h_{n+1} & 0 \\
        \end{array}
      \right)
\end{align}
and hence stationary AKNS hierarchy can be introduced as follows
\begin{equation}
  \textrm{s-AKNS}_{n+1}(q,p)=-\left(
                    \begin{array}{c}
                      h_{n+1}(q,p) \\
                      f_{n+1}(q,p) \\
                    \end{array}
                  \right)=0,~~n\in\mathbb{N}_0.
\end{equation}
Explicitly,
\begin{align}
  \textrm{s-AKNS}_1(q,p)=&~\left(
                    \begin{array}{c}
                      -p_x+c_1(-p) \\
                      -q_x+c_1q \\
                    \end{array}
                  \right)=0,\nonumber\\
\textrm{s-AKNS}_2(q,p)=&~\left(
                    \begin{array}{c}
                      -p_{xx}+2p^2q+c_1(-p_x)+c_2(-p) \\
                     q_{xx}-2pq^2+c_1(-q_x)+c_2q \\
                    \end{array}
                  \right)=0,\\
\textrm{s-AKNS}_3(q,p)=&~\left(
                    \begin{array}{c}
                   -p_{xxx}+6pp_xq+c_1(-p_{xx}+2p^2q)+c_2(-p_x)+c_3(-p) \\
                   -q_{xxx}+6pqq_x+c_1(q_{xx}-2pq^2)+c_2(-q_x)+c_3 q \\
                    \end{array}
                  \right)=0,~~\textrm{etc.}\nonumber
\end{align}
Similarly, from (\ref{2.4}) it follows
\begin{align}
 0=&~U_{t_{n+1}}-V_{n+1,x}+[U(z),V_{n+1}(z)]\nonumber\\
 =&~\left(
      \begin{array}{cc}
        0 & q_{t_{n+1}} -F_{n,x}-zF_n+2qG_{n+1} \\
        p_{t_{n+1}}+H_{n,x}-zH_n-2pG_{n+1}& 0 \\
      \end{array}
    \right)\nonumber\\
    =&\left(
        \begin{array}{cc}
          0 & q_{t_{n+1}}-f_{n+1} \\
          p_{t_{n+1}}-h_{n+1} & 0 \\
        \end{array}
      \right).
\end{align}
Thus, one finally derives time-dependent AKNS hierarchy
\begin{equation}
  \textrm{AKNS}_{n+1}(q,p)=\left(
                    \begin{array}{c}
                      p_{t_{n+1}}-h_{n+1}(q,p) \\
                      q_{t_{n+1}}-f_{n+1}(q,p) \\
                    \end{array}
                  \right)=0,~~n\in\mathbb{N}_0.
\end{equation}
Explicitly,
\begin{align}
  \textrm{AKNS}_1(q,p)=&~\left(
                    \begin{array}{c}
                      p_{t_1}- p_x+c_1(-p) \\
                      q_{t_1}- q_x+c_1q \\
                    \end{array}
                  \right)=0,\nonumber\\
\textrm{AKNS}_2(q,p)=&~\left(
                    \begin{array}{c}
                     p_{t_2}-p_{xx}+2p^2q+c_1(-p_x)+c_2(-p) \\
                     q_{t_2}+q_{xx}-2pq^2+c_1(-q_x)+c_2 q \\
                    \end{array}
                  \right)=0,\\
\textrm{AKNS}_3(q,p)=&~\left(
                    \begin{array}{c}
                   p_{t_3}-  p_{xxx}+6pp_xq+c_1(-p_{xx}+2p^2q)+c_2(-p_x)+c_3(-p) \\
                   q_{t_3}- q_{xxx}+6pqq_x+c_1(q_{xx}-2pq^2)+c_2(-q_x)+c_3 q \\
                    \end{array}
                  \right)=0,~~\textrm{etc.}\nonumber
\end{align}

Next we turn to discuss the relations between KP and AKNS hierarchy.
To this end we denote by
\begin{equation}
  \widehat{\textrm{AKNS}}_{n+1}(q,p)=\textrm{AKNS}_{n+1}(q,p)|_{c_k=0,k=1,\ldots,n+1},~~n\in\mathbb{N}_0,
\end{equation}
the corresponding homogeneous AKNS equations.

\newtheorem{thmnew}{Theorem}[section]
\begin{thmnew}[see \cite{Sidorenkot1991} or \cite{cheng1992}]
Assume $p,q$ is a compatible solution of the system
 \begin{align}
    &\widehat{\textrm{AKNS}}_{2}(q,p)=0, \label{st2.38}\\
    &\widehat{\textrm{AKNS}}_{r+1}(q,p)=0, ~~r\geq 2.\label{st2.39}
 \end{align}
Then
 $$u(x,y,t_{r+1})=-q(x, -y,(-1)^{r}t_{r+1})p(x, -y,(-1)^{r}t_{r+1})$$
gives a solution of the $n$th KP equation.
\end{thmnew}

Given these preparations, we turn to study the KP equations.
Now $q,p$ are considered as functions of variables $t_1,t_2,t_{r+1}$ with $t_1=x,t_2=y$ and
we shall start from the following auxiliary linear problem \footnote{One can also start from the
following linear problems
 \begin{equation}\label{2.143433}
\begin{split}
  &\psi_x(z)=U(z)\psi(z), \\
  &\psi_{t_{m+1}}(z)=\widehat{V}_{m+1}(z)\psi(z),~~z\in\mathbb{C},~~m=2,3,\ldots
  \end{split}
\end{equation}
where $q,p$ are considered as functions of $x,t_2,t_3,\ldots$
and obtain similar results.
}
\begin{equation}\label{2.14k}
\begin{split}
  &\psi_x(z)=U(z)\psi(z), \\
  &\psi_y(z)=\widehat{V}_2(z)\psi(z),\\
  &\psi_{t_{r+1}}(z)=\widehat{V}_{r+1}(z)\psi(z),~~z\in\mathbb{C},~~r\geq 2,
  \end{split}
\end{equation}
where $\widehat{V}_k(z)=V_{k}(z)|_{c_{\ell}=0,\ell=1,\ldots,k},$
and
$\psi(z)=(\psi_1(z,x,y,t_{r+1}),\psi_2(z,x,y,t_{r+1}))^T.$
Let
$$\psi^\pm(z)=(\psi_1^{\pm}(z,x,y,t_{r+1}),\psi_2^{\pm}(z,x,y,t_{r+1}))$$
 be two fundamental solutions of linear system (\ref{2.14k}). Then we can define three squared basis functions $\mathscr{G}, \mathscr{F}, \mathscr{H}$ by
\begin{equation}\label{2.6}
\begin{split}
  \mathscr{G}(z)=&~ \frac{1}{2}(\psi_1^+(z)\psi_2^-(z)+\psi_1^-(z)\psi_2^+(z)), \\
  \mathscr{F}(z)=&~\psi_1^+(z)\psi_1^-(z),\\
  \mathscr{H}(z)=&~\psi_2^+(z)\psi_2^-(z).
\end{split}
\end{equation}
Using
(\ref{2.1}), (\ref{2.2}) and (\ref{2.14k}), one finds
$\mathscr{G}, \mathscr{F},$ and $\mathscr{H},$ satisfy the following linear system
\begin{align}
 & \mathscr{G}_{x}=p \mathscr{F}+q\mathscr{H},\label{2.16k} \\
 & \mathscr{G}_{y}=\widehat{F}_{1}\mathscr{H}-\widehat{H}_1 \mathscr{F},\\
 & \mathscr{G}_{t_{r+1}}=\widehat{F}_{r}\mathscr{H}-\widehat{H}_r \mathscr{F},\\
 & \mathscr{F}_{x}=2q\mathscr{G}-z\mathscr{F},\label{2.19k}\\
 &\mathscr{F}_{y}=2\widehat{F}_1\mathscr{G}-2\widehat{G}_2\mathscr{F},\label{2.20kp}\\
 &\mathscr{F}_{t_{r+1}}=2\widehat{F}_r\mathscr{G}-2\widehat{G}_{r+1}\mathscr{F},\label{2.21kp}\\
 &\mathscr{H}_{x}=2p\mathscr{G}+z\mathscr{H},\label{2.22k}\\
 &\mathscr{H}_{y}=2\widehat{G}_2\mathscr{H}-2\widehat{H}_1\mathscr{G},\label{2.23k}\\
 & \mathscr{H}_{t_{r+1}}=2\widehat{G}_{r+1}\mathscr{H}-2\widehat{H}_r\mathscr{G}.\label{2.24k}
\end{align}
For fixed $r$, solutions of linear system (\ref{2.16k})-(\ref{2.24k}) are connected with the following
basic initial value problem of AKNS system (\ref{st2.51}).

\newtheorem{th1}{Theorem}[section]
\begin{th1}
Assume $q,p\in C^\infty(\mathbb{R}^{2+1})$.
Moreover, suppose $q,p$ is a solution satisfying (\ref{st2.38}), (\ref{st2.39}).
Then the collection of polynomials
\begin{equation}\label{2.29k}
\{(\mathscr{G},\mathscr{F},\mathscr{H})|~
(G_{j+1},F_{j},H_{j}),~ j\in\mathbb{N}_0\}
\end{equation}
gives a sequence of special solutions for (\ref{2.16k})-(\ref{2.24k}). In particular, $(G_{n+1},$ $F_{n},H_{n})$,$ n\in\mathbb{N}_0,$
corresponds to solutions of the following initial problem
 \begin{align}\label{st2.51}
\begin{cases}
 q_{y}-\hat{f}_{2}=0,\\
 p_{y}-\hat{h}_{2}=0,\\
 q_{t_{r+1}}-\hat{f}_{r+1}=0,\\
 p_{t_{r+1}}-\hat{h}_{r+1}=0,\\
 -f_{n+1}=0,\\
 -h_{n+1}=0,~~ r\geq 2.
 \end{cases}
\end{align}

\end{th1}
\proof
From (\ref{2.3}) and (\ref{2.4}) we know
(\ref{st2.51}) is equivalent to the following zero
curvature representation
\begin{equation}\label{2.9k}
\begin{split}
  &\frac{\partial}{\partial x}V_j(z)-\frac{\partial}{\partial t_j}U(z)=[U(z),V_j(z)],~j=2,r+1,n+1,
\end{split}
\end{equation}
introducing $\frac{\partial}{\partial t_{n+1}}=0.$
Thus, it suffices to show
\begin{equation}\label{2.30k}
\begin{split}
  [V_{j+1}(z)-\frac{\partial}{\partial t_j},V_{k+1}(z)-\frac{\partial}{\partial t_k}]
  &=0,~~j,~k=2,r+1,n+1.
  \end{split}
\end{equation}
and we proceed to prove this as follows. Define $2\times 2$
matrix-valued differential expression
 \begin{align}
   L=&\frac{1}{2}\left(
                  \begin{array}{cc}
                    \frac{d}{dx} & -q \\
                    p &  \frac{d}{dx} \\
                  \end{array}
                \right),
                \\
P_{n+1}=&\sum_{j=0}^{n+1}\left(
                           \begin{array}{cc}
                             -g_{n+1-\ell} & f_{n-\ell} \\
                             -h_{n-\ell} &  g_{n+1-\ell}\\
                           \end{array}
                         \right)L^\ell,~ n\in\mathbb{N}_0,~f_{-1}=h_{-1}=0.
 \end{align}
Then it is not difficult to verify that (\ref{2.9k}) is equivalent to
\begin{equation*}
  [P_{j+1}-\frac{\partial}{\partial t_j},L]=0,~~j=2, r+1, n+1.
\end{equation*}
By Corollary 2 of Theorem 4.2 in \cite{Krichever} it follows
 \begin{equation*}
  [P_{j+1}-\frac{\partial}{\partial t_j},P_{k+1}-\frac{\partial}{\partial t_k}]=0,~~j,~k=2, r+1, n+1.
\end{equation*}
and hence we obtain
\begin{align}
  [V_{j+1}(z)-\frac{\partial}{\partial t_j},V_{k+1}(z)-\frac{\partial}{\partial t_k}]&=[P_{j+1}-\frac{\partial}{\partial t_j},P_{k+1}-\frac{\partial}{\partial t_k}]|_{\textrm{ker}(M-z)}\nonumber\\
  &=0,~~j,~k=1,\ldots,m,
\end{align}
where
\begin{equation*}
  \textrm{ker}(M-z)=\{\Psi=\left(
                             \begin{array}{c}
                               \psi_1 \\
                               \psi_2 \\
                             \end{array}
                           \right):\mathbb{R}^{s+1}\rightarrow \mathbb{C}_{\infty}^2|(M-z)\Psi=0
  \},~z\in\mathbb{C}.
\end{equation*}
 This completes the proof.\qed

\newtheorem{rem1}[th1]{Remark}
\begin{rem1}
This proposition can also be proved by introducing a fundamental meromorphic function $\phi$
on Riemann surface $X$, where
$q,p$ should be considered as functions of $x,y,t_{r+1}$ (see \cite{gesztesy1998}).

\end{rem1}
\section{Spectral curve, Dubrovin-type equations and trace formula}
In what follows, we shall introduce the spectral curve associated with
KP hierarchy. To emphasize the difference between different solutions in (\ref{2.29k}),
we add a subscript in each of $\mathscr{G},\mathscr{F},\mathscr{H}$, that is,
\begin{equation*}
(\mathscr{G}_{n+1},\mathscr{F}_{n},\mathscr{H}_{n})=(G_{n+1},F_{n},H_{n}),~n\in\mathbb{N}. \end{equation*}
 Then using (\ref{2.16k})-(\ref{2.24k}) and theorem 2.1, we have
\begin{equation}
(\mathscr{G}^2_{n+1}-\mathscr{F}_{n}\mathscr{H}_{n})_{t_j}=0,~~n\in\mathbb{N},~j=1,2,t_{r+1},
\end{equation}
and hence $\mathscr{G}^2_{n+1}-\mathscr{F}_{n}\mathscr{H}_{n}$ is $x,y,t_{r+1}$-independent implying
\begin{equation}\label{3.3zp}
\mathscr{G}^2_{n+1}-\mathscr{F}_{n}\mathscr{H}_{n}=\mathscr{R}_{2n+2},~~n\in\mathbb{N},
\end{equation}
where integration constant $\mathscr{R}_{2n+2}$ is a polynomial of degree $2n+2.$ Let
\begin{equation}\label{2.21}
\begin{split}
  \mathscr{R}_{2n+2}&(z)=\sum_{j=0}^{2n+2-j}(-1)^js_j z^{2n+2-j},\\
  &s_0=1/4,~s_{j}\in\mathbb{C},j=1,\ldots,2n,
\end{split}
\end{equation}
and hyperelliptic curve associated with the $r$th KP equation is then introduced
as follows:
\begin{equation}\label{2.22}
  X: \mathscr{P}(z,y)=y^2-4\mathscr{R}_{2n+2}(z)=y^2-4\mathscr{G}_{n+1}^2+4\mathscr{F}_n\mathscr{H}_n=0.
\end{equation}
The curve $X$ is compactified by joining two points $P_{\infty\pm}$ at infinity but for notational simplicity the compactification is also denoted by $X$. Points $P$ on $X\backslash \{P_{\infty\pm}\}$ are denote by pairs $(z,y)$, where $y(\cdot)$ is the
meromorphic function on $X$ satisfying $\mathscr{P}(z,y)=0.$ The complex structure
on $X$ is then defined in the usual way. Hence, $X$ becomes
a two-sheeted hyperelliptic Riemann surface of (arithmetic) genus $n$ in a standard manner.
Moreover, we denote the upper and lower sheets $\Pi_\pm$ by
\begin{equation*}
  \Pi_\pm=\{(z,\pm2\sqrt{\mathscr{R}_{2n+2}(z)})\in X|z\in \Pi\},
\end{equation*}
where $\Pi$ denotes the cut plane $\mathbb{C}\backslash\mathcal{C}$
and $\mathcal{C}$ is the union of $n$ nonintersecting cuts joining two different
branches of $\sqrt{\mathscr{R}_{2n+2}(z)}$. The holomorphic sheet exchange map on $X$ is defined by
\begin{align*}
&*:\quad X\rightarrow X, \\
&P=(z,2\sqrt{\mathscr{R}_{2n+2}(z)})\mapsto P^{*}=(z,-2\sqrt{\mathscr{R}_{2n+2}(z)}),\quad\\
&P_{\infty\pm}\mapsto P_{\infty\pm}^{*}=P_{\infty\mp}.
\end{align*}
 Moreover, positive divisors on $X$ of degree $n$
   are denoted by
        \begin{equation}\label{3.4xyz}
          \mathcal{D}_{P_1,\ldots,P_{r-2}}:
             \begin{cases}
              X\rightarrow \mathbb{N}_0,\\
              P\rightarrow \mathcal{D}_{P_1,\ldots,P_{n}}(P)=
                \begin{cases}
                  \textrm{ $k$ if $P$ occurs $k$
                      times in $\{P_1,\ldots,P_{n}\},$}\\
                   \textrm{ $0$ if $P \notin
                     $$ \{P_1,\ldots,P_{n}\}.$}
                \end{cases}
             \end{cases}
        \end{equation}
        In particular, the divisor $\left(\phi(\cdot)\right)$ of a meromorphic function
        $\phi(\cdot)$ on X is defined by
        \begin{eqnarray}\label{3.4a0xyz}
          \left(\phi(\cdot)\right):
          X \rightarrow \mathbb{Z},\quad
          P\mapsto \omega_{\phi}(P),
          \end{eqnarray}
        where $\omega_{\phi}(P)=m_0\in\mathbb{Z}$ if
        $(\phi\circ\zeta_{P}^{-1})(\zeta)=\sum_{n=m_0}^{\infty}c_n(P)\zeta^n$
        for some $m_0\in\mathbb{Z}$ by using a chart $(U_{P}, \zeta_P)$ near $P\in X.$
        Finally, we introduce symmetric functions of $x_1,\ldots,x_n$
\begin{equation}\label{2.25}
\begin{split}
& \Psi_{0}(\underline{x})=1,~~\Psi_{1}(\underline{x})=\sum_{j=1}^{n}x_j,~~\Psi_{2}(\underline{x})
=\sum_{ j,k=1,j<k
}^{n}x_jx_k, \\
&\Psi_{3}(\underline{x})
=\sum\nolimits_{  j,k,\ell=1,j<k<\ell
 }^{n}x_jx_kx_\ell, ~~\textrm{etc.,}
 \end{split}
\end{equation}
 where
 $$\underline{x}=(x_1,\ldots,x_{n}).$$

Now we turn to the parameter representations of $\mathscr{G}_{n+1}, \mathscr{F}_{n}, \mathscr{H}_{n}$,
which are described by the evolution of auxiliary spectrum points $\mu_j(x,y,t_{r+1}),$ \linebreak $\nu_j(x,y,t_{r+1}),j=1,\ldots,n$. This procedure is standard, which is similar with 1+1 dimensional case.

\newtheorem{th2}{Theorem}[section]
\begin{th2}
Solutions of (\ref{2.16k})-(\ref{2.24k}) can also be expressed as
\begin{equation}\label{2.23}
  \begin{split}
  &\mathscr{G}_{n+1}=\sum_{j=0}^{n+1} (-1)^j \dot{g}_j(x,y,t_{r+1})z^{n+1-j},~~ \\
 & \mathscr{F}_{n}=-q(x,y,t_{r+1})\prod_{j=1}^{n} (z-\mu_j(x,y,t_{r+1})),\\
 & \mathscr{H}_{n}=p(x,y,t_{r+1})\prod_{j=1}^{n} (z-\nu_j(x,y,t_{r+1})),
  \end{split}
\end{equation}
where
\begin{align}\label{2.24}
  \dot{g}_0=&~1/2,~\dot{g}_1=s_1,~\nonumber\\
  \dot{g}_2=&~-pq-s_1^2+s_2,\\
  \dot{g}_j=&~-\sum_{\nu=1}^{j-1}\dot{g}_\nu \dot{g}_{j-\nu}-pq \sum_{\alpha+\beta=j}\Psi_{\alpha}(\underline{\mu})\Psi_{j-\beta}(\underline{\nu})+s_j,~s=3,\ldots,n+1,\nonumber
\end{align}
and $\{\mu_j(x,y,t_{r+1})\}_{j=1}^n, \{\nu_j(x,y,t_{r+1})\}_{j=1}^n$ are $n$ roots of $\mathscr{F}_{n},\mathscr{H}_{n}$, respectively. In particular, if $\{\mu_j(x,y,t_{r+1})\}_{j=1}^n$ are mutually distinct and finite, then
 they satisfy the
 Dubrovin-type equations
 \begin{align}
    &\mu_{j,x}=-\frac{2\sqrt{\mathscr{R}_{2n+2}(\mu_j)}}{ \prod_{k\neq j}(\mu_j-\mu_k)},\label{2.28l}\\
    &\mu_{j,y}=-\frac{2\widehat{F}_1(\mu_j)\sqrt{\mathscr{R}_{2n+2}(\mu_j)}}{q \prod_{k\neq j}(\mu_j-\mu_k)},\label{2.29m}\\
    &\mu_{j,t_{r+1}}=-\frac{2\widehat{F}_{r}(\mu_j)\sqrt{\mathscr{R}_{2n+2}(\mu_j)}}{q \prod_{k\neq j}(\mu_j-\mu_k)},\label{2.29n}
 \end{align}
 and similar statement is also true for  $\{\nu_j(x,y,t_{r+1})\}_{j=1}^n$, where (\ref{2.28l})-(\ref{2.29n})
 change to
 \begin{align}
    &\nu_{j,x}=-\frac{2\sqrt{\mathscr{R}_{2n+2}(\nu_j)}}{ \prod_{k\neq j}(\nu_j-\nu_k)},\label{2.30}\\
    &\nu_{j,y}=-\frac{2\widehat{H}_1(\mu_j)\sqrt{\mathscr{R}_{2n+2}(\mu_j)}}{p \prod_{k\neq j}(\mu_j-\mu_k)},\label{2.30m}\\
    &\nu_{j,t_{r+1}}=-\frac{2\widehat{H}_2(\mu_j)\sqrt{\mathscr{R}_{2n+2}(\mu_j)}}{p \prod_{k\neq j}(\mu_j-\mu_k)}.\label{2.30n}
 \end{align}
Finally, $q,p$ and $\mu_j,\nu_j$ are connected by the following trace formula
 \begin{equation}\label{2.32}
   \sum_{j=1}^{n}\mu_j=\frac{q_x}{q}-2s_1,~~\sum_{j=1}^{n}\nu_j=-\frac{p_x}{p}+2s_1.
 \end{equation}
\end{th2}
\proof First, insertion of (\ref{2.23}) into (\ref{3.3zp}), (\ref{2.21}) and a comparison powers of $z$ yields (\ref{2.24}). Then by (\ref{2.19k}), (\ref{2.20kp}), (\ref{2.21kp}), (\ref{2.22k}), (\ref{2.23k})
and (\ref{2.24k}),
taking into account (\ref{2.22}), (\ref{2.23}),
one derives (\ref{2.28l})-(\ref{2.30n}). Moreover, combining (\ref{st2.23}) with (\ref{2.24}) yields relations
\begin{equation}\label{2.26}
\begin{split}
  &\dot{g}_j=(-1)^j g_j,~~
   c_1=-2 s_1,~~c_2=2(s_2-s_1^2),\\
  &c_3= 4s_1(s_2-s_1^2)-2s_3,~~\textrm{etc.}
\end{split}
\end{equation}
and formula (\ref{2.32}) is the direct result of (\ref{2.23}), (\ref{2.26}) and theorem 2.1.
\qed

\section{Baker-Akhiezer function}
Baker-Akhiezer function plays a very important role in finite gap integration of
soliton equations and it permits us to obtain the Riemann theta function representation for solutions of a
given equation and there are numerous articles have been
devoted to this subject \cite{Belokolos1994},~\cite{Kamchatnov},~\cite{gesztesy1999},~\cite{gesztesy2003a}.
In this section, we shall construct the explicit form of "Baker-Akhiezer function" associated with the $r$th KP equation, which consists of spectral parameter $z$, potentials $q,p$,
and then study its analytic properties.
Moreover, we will find the conservation relations of soliton equations play a key role in the construction of Baker-Akhiezer function, which reflects symmetry is the intrinsic character of classical integrable system.

To find explicit form of Baker-Akhiezer function
$$\psi(P)=(\psi_1(P,x,y,t_{r+1}),\psi_2(P,x,y,t_{r+1})),~P\in X,$$
which satisfies
\begin{equation}\label{4.1k}
\begin{split}
  &\psi_x(P)=~U(z)\psi(P),~
  \psi_y(P)=\widehat{V}_2(z)\psi(P),~\psi_{t_{r+1}}(P)=\widehat{V}_{r+1}(z)\psi(P),\\
  &\psi_1(P,x_0,y_0,t_{r+1,0})=1,~~(P,x_0,y_0,t_{r+1,0})\in X\times\mathbb{R}^3,
  \end{split}
\end{equation}
we need some preparations.

\newtheorem{lem4.1}{Lemma}[section]
\begin{lem4.1}
Suppose $q,p\in C^{\infty}(\mathbb{R}^{2+1})$ and $z\in\mathbb{C}$.
Then
$\mathscr{G}_{n+1}, \mathscr{F}_{n},
\mathscr{H}_n$ and basic fundamental solutions $(\psi_1^\pm,\psi_2^\pm)$ of linear system (\ref{4.1k})
have the following algebraic relation:
\begin{align}
 &(\psi_1^+(z)\psi_2^-(z)-\psi_1^-(z)\psi_2^+(z))^2=4 \mathscr{R}_{2n+2}^2(z).\label{4.2k}
 \end{align}
 If we take $\psi_1^+(z)\psi_2^-(z)-\psi_1^-(z)\psi_2^+(z)=2\sqrt{\mathscr{R}_{2n+2}(z)},$
 then
 \begin{align}
 &\psi_1^+(z)\psi_2^-(z)=\mathscr{G}_{n+1}(z)+\sqrt{\mathscr{R}_{2n+2}(z)},\\
 &\psi_1^-(z)\psi_2^+(z)=\mathscr{G}_{n+1}(z)-\sqrt{\mathscr{R}_{2n+2}(z)},\\
 &\frac{\psi_2^\pm(z)}{\psi_1^\pm(z)}=\frac{\mathscr{H}_n(z)}{\mathscr{G}_{n+1}(z)\pm \sqrt{\mathscr{R}_{2n+2}(z)}}=\frac{\mathscr{G}_{n+1}\mp \sqrt{\mathscr{R}_{2n+2}(z)}}{\mathscr{F}_n}.\label{4.5k}
\end{align}
\end{lem4.1}
\proof Expressions (\ref{4.2k})-(\ref{4.5k}) can easily be verified by (\ref{2.6}), (\ref{3.3zp}). \qed

\newtheorem{lem4.2}[lem4.1]{Lemma}
\begin{lem4.2}
Suppose $q,p\in C^{\infty}(\mathbb{R}^{2+1})$. Then we have the following relations
\begin{align}
 &\big[\frac{q(x,y,t_{r+1})}{\mathscr{F}_n(z,x,y,t_{r+1})}\big]_{t_{j+1}}=\big[\frac{{\widehat{F}_j(z,x,y,t_{r+1})}}{\mathscr{F}_n(z,x,y,t_{r+1})}\big]_{x},\\
 &
 \big[\frac{p(x,y,t_{r+1})}{\mathscr{H}_n(z,x,y,t_{r+1})}\big]_{t_{j+1}}=-\big[\frac{\widehat{H}_j(z,x,y,t_{r+1})}{\mathscr{F}_n(z,x,y,t_{r+1})}\big]_{x},~~j=1,r,\label{4.6kp}\\
 &\big[\frac{{\widehat{F}_1(z,x,y,t_{r+1})}}{\mathscr{F}_n(z,x,y,t_{r+1})}\big]_{t_{r+1}}=\big[\frac{{\widehat{F}_2(z,x,y,t_{r+1})}}{\mathscr{F}_n(z,x,y,t_{r+1})}\big]_{y}, \\
  &\big[\frac{{\widehat{H}_1(z,x,y,t_{r+1})}}{\mathscr{F}_n(z,x,y,t_{r+1})}\big]_{t_{r+1}}
  =\big[\frac{{\widehat{H}_2(z,x,y,t_{r+1})}}{\mathscr{F}_n(z,x,y,t_{r+1})}\big]_{y}.\label{4.8kp}
\end{align}
\end{lem4.2}
\proof The proof is straightforward. By (\ref{2.20kp}), (\ref{2.21kp}) and
\begin{equation}
  q_{t_{j+1}}-\widehat{F}_{j,x}-z\widehat{F}_j+2q\widehat{G}_{j+1}=0,~~j=1,r,
\end{equation}
we have
\begin{equation}
\begin{split}
  \Big[\frac{q}{\mathscr{F}_n}\Big]_{t_j}=&~\frac{q_{t_j}}{\mathscr{F}_n}-
  \frac{q \mathscr{F}_{n,t_j}}{\mathscr{F}_n^2}\nonumber\\
  =&~\frac{\widehat{F}_{j,x}+z\widehat{F}_j-2q\widehat{G}_{j+1}}{\mathscr{F}_n}-\frac{q(2\widehat{F}_j\mathscr{G}_{n+1}-2\widehat{G}_{j+1}
  \mathscr{F}_n)}{\mathscr{F}_n^2}\nonumber\\
  =&\frac{\widehat{F}_{j,x}+z\widehat{F}_j\mathscr{F}_n-2q\widehat{F}_j\mathscr{G}_{n+1}}{\mathscr{F}_n^2}\nonumber\\
  =&\frac{\widehat{F}_{j,x}-\widehat{F}_j\mathscr{F}_{n,x}}{\mathscr{F}_n^2}= \big[\frac{{\widehat{F}_j}}{\mathscr{F}_n}\big]_{x},~~j=1,r.\nonumber\\
  \end{split}
\end{equation}
and (\ref{4.6kp})-(\ref{4.8kp}) follows in a similar manner. \qed\vspace{0.3cm}

Now we lift the fundamental solution $(\psi_1^\pm(z),\psi_2^\pm(z))$, $z\in\mathbb{C},$ to Riemann surface $X$.
Define the Baker-Akhiezer function by
\begin{align*}
&\psi_j(P,x,y,t_{r+1})=  \psi_j^+(z,x,y,t_{r+1}),~~ j=1,2,~~ y(P)=2\sqrt{\mathscr{R}_{2n+2}(z)}, ~~\textrm{for}~~ P\in \Pi_+, \\
& \psi_j(P,x,y,t_{r+1})= \psi_j^-(z,x,y,t_{r+1}), ~~j=1,2, ~~y(P)=-2\sqrt{\mathscr{R}_{2n+2}(z)}, ~~\textrm{for}~~ P\in \Pi_-, \\
&\lim_{P\rightarrow P_0}\psi_j(P,x,t_{r+1})= \lim_{P\rightarrow P_0}\psi_j(P^*,x,t_{r+1}),~~\textrm{for}~~j=1,2,~~P\in \Pi_\pm,~~P_0\in\mathcal{C},
\end{align*}
where we choose the branches of $\sqrt{\mathscr{R}_{2n+2}(\cdot)}$ satisfying
$$\lim_{z\rightarrow\infty}\frac{\mathscr{G}_{n+1}}{\sqrt{\mathscr{R}_{2n+2}(z)}}=1.$$
Moreover,
let
\begin{align*}
  \hat{\mu}_j(x,y,t)=&\Big(\mu_j(x,y,t_{r+1}), -2\sqrt{\mathscr{R}_{2n+2}(\mu_j(x,y,t_{r+1}))}\Big)\in \Pi_-,\\
   \hat{\nu}_j(x,y,t)=&\Big(\nu_j(x,y,t_{r+1}), 2\sqrt{\mathscr{R}_{2n+2}(\mu_j(x,y,t_{r+1}))}\Big)\in \Pi_+,~~j=1,\ldots, n.
\end{align*}
Based on above preparations, we shall study explicit forms of Baker-Akhiezer function $\psi_j(P),j=1,2$.
\newtheorem{thm4.2}[lem4.1]{Theorem}
\begin{thm4.2}
Suppose $q,p\in C^{\infty}(\mathbb{R}^{2+1})$.
Then Baker-Akhiezer function satisfying condition (\ref{4.1k}) can be expressed as
 \begin{align}
 \psi_1(P,x,y,t_{r+1})=&~\sqrt{\frac{\mathscr{F}_n(z,x,y,t_{r+1})}{\mathscr{F}_n(z,x_0,y_0,t_{r+1,0})}}\exp\Big(-\frac{1}{2} \int_{x_0}^x\frac{q(x^\prime,y,t_{r+1})y(P)}{\mathscr{F}_n(z,x^\prime,y,t_{r+1})}dx^\prime \nonumber \\
&-  \frac{y(P)}{2}\int_{y_0}^y\frac{\widehat{F}_1(z,x_0,y^\prime,t_{r+1})}{\mathscr{F}_n(z,x_0,y^\prime,t_{r+1})}dy^\prime-\frac{y(P)}{2}
\int_{t_{r+1,0}}^{t_{r+1}}\frac{\widehat{F}_r(z,x_0,y_0,t^\prime)}{\mathscr{F}_n(z,x_0,y_0,t^\prime)}dt^\prime\Big),
\label{4.10kp}\\
\psi_2(P,x,y,t_{r+1})=&~\sqrt{\frac{\mathscr{H}_n(z,x,y,t_{r+1})}{\mathscr{F}_n(z,x_0,y_0,t_{r+1,0})}}\exp\Big(\frac{1}{2} \int_{x_0}^x\frac{p(x^\prime,y,t_{r+1})y(P)}{\mathscr{F}_n(z,x^\prime,y,t)}dx^\prime \nonumber \\
&-\frac{y(P)}{2}\int_{y_0}^y\frac{\widehat{H}_1(z,x_0,y^\prime,t)}{\mathscr{F}_n(z,x_0,y^\prime,t)}dy^\prime- \frac{y(P)}{2}\int_{t_{r+1,0}}^{t_{r+1}}\frac{\widehat{H}_r(z,x_0,y_0,t^\prime)}{\mathscr{F}_n(z,x_0,y_0,t^\prime)}dt^\prime\Big).
\label{4.11kp}
 \end{align}
\end{thm4.2}
\proof
By (\ref{2.14k}), (\ref{2.19k})-(\ref{2.24k}), (\ref{4.5k}), one obtains
\begin{align}
  \psi_{1,x}^\pm(z)=&-\frac{z}{2}\psi_1^\pm(z)+q\psi_2^\pm(z)\nonumber\\
  =&\left[\left(-\frac{q\mathscr{G}_{n+1}(z)}{\mathscr{F}_{n}(z)}+\frac{1}{2}\frac{\mathscr{F}_{n,x}(z)}{\mathscr{F}_n(z)}\right)
  +q\frac{\psi_2^\pm(z)}{\psi_1^\pm(z)}\right]\psi_1^\pm(z)\nonumber\\
  =&\left[\left(-\frac{q\mathscr{G}_{n+1}(z)}{\mathscr{F}_{n}(z)}+\frac{1}{2}\frac{\mathscr{F}_{n,x}(z)}{\mathscr{F}_n(z)}\right)
  +q \frac{\mathscr{H}_n(z)}{\mathscr{G}_{n+1}(z)\pm\sqrt{\mathscr{R}_{2n+2}(z)}}\right]\psi_1^\pm(z)\nonumber\\
  =& \frac{\mp q \sqrt{\mathscr{R}_{2n+2}(z)}+\frac{1}{2}\mathscr{F}_{n,x}(z)}{\mathscr{F}_{n}(z)}\psi_1^\pm(z),\\
   \psi_{1,t_{j+1}}^\pm(z)=&-\widehat{G}_{j+1}(z)\psi_1^\pm(z)+\widehat{F}_{j}(z)\psi_2^\pm(z)\nonumber\\
  =&\left[\left(-\frac{\widehat{F}_j(z)\mathscr{G}_{n+1}(z)}{\mathscr{F}_{n}(z)}
  +\frac{1}{2}\frac{\mathscr{F}_{n,t_{j+1}}(z)}{\mathscr{F}_n(z)}\right)
  +\widehat{F}_{j}(z)\frac{\psi_2^\pm(z)}{\psi_1^\pm(z)}\right]\psi_1^\pm(z)\nonumber\\
  =&\left[\left(-\frac{\widehat{F}_j(z)\mathscr{G}_{n+1}(z)}{\mathscr{F}_{n}(z)}
  +\frac{1}{2}\frac{\mathscr{F}_{n,t_{j+1}(z)}}{\mathscr{F}_n(z)}\right)
  +\widehat{F}_j(z) \frac{\mathscr{H}_n(z)}{\mathscr{G}_{n+1}(z)\pm\sqrt{\mathscr{R}_{2n+2}(z)}}\right]\psi_1^\pm(z)\nonumber\\
  =& \frac{\mp \widehat{F}_j(z) \sqrt{\mathscr{R}_{2n+2}(z)}+\frac{1}{2}\mathscr{F}_{n,t_{j+1}}(z)}{\mathscr{F}_{n}(z)}\psi_1^\pm(z),~~j=1,r,
  \end{align}
  and similarly
  \begin{align}
    \psi_{2,x}^\pm(z)=&~p\psi_1^\pm(z)+\frac{z}{2}\psi_2^\pm(z)\nonumber\\
  =&~\left[p\frac{\psi_1^\pm(z)}{\psi_2^\pm(z)}
  +\left(\frac{\mathscr{H}_{n,x}(z)}{2\mathscr{H}_n(z)}-\frac{p\mathscr{G}_{n+1}(z)}{\mathscr{H}_n(z)}\right)\right]\psi_2^\pm(z)\nonumber\\
  =&~\left[p\frac{\mathscr{G}_{n+1}(z)\pm \sqrt{\mathscr{R}_{2n+2}(z)}}{\mathscr{H}_{n}(z)}+\left(\frac{\mathscr{H}_{n,x}(z)}{2\mathscr{H}_n(z)}-\frac{p\mathscr{G}_{n+1}(z)}{\mathscr{H}_n(z)}\right)\right]\psi_2^\pm(z)\nonumber\\
  =&~\frac{\pm p \sqrt{\mathscr{R}_{2n+2}(z)}+\frac{1}{2}\mathscr{H}_{n,x}(z)}{\mathscr{H}_{n}(z)}\psi_2^\pm(z),\\
  \psi_{2,t_{j+1}}^\pm(z)=&~-\widehat{H}_j(z)\psi_1^\pm(z)+\widehat{G}_{j+1}(z)\psi_2^\pm(z)\nonumber\\
  =&~\left[-\widehat{H}_j(z)\frac{\psi_1^\pm(z)}{\psi_2^\pm(z)}+
  \left(\frac{\mathscr{H}_{n,t_{j+1}}(z)}{2\mathscr{H}_n(z)}+\widehat{H}_j(z)\frac{\mathscr{G}_{n+1}(z)}{\mathscr{H}_n(z)}\right)\right]\psi_2^\pm(z)\nonumber\\
  =&~\left[-\widehat{H}_j(z)\frac{\mathscr{G}_{n+1}(z)\pm \sqrt{\mathscr{R}_{2n+2}(z)}}{\mathscr{H}_{n}(z)}+\left(\frac{\mathscr{H}_{n,t_{j+1}}(z)}{2\mathscr{H}_n(z)}
  +\widehat{H}_j(z)\frac{\mathscr{G}_{n+1}(z)}{\mathscr{H}_n(z)}\right)\right]\psi_2^\pm(z)\nonumber\\
  =&~\frac{\mp \widehat{H}_j(z) \sqrt{\mathscr{R}_{2n+2}(z)}+\frac{1}{2}\mathscr{H}_{n,t_{j+1}}(z)}{\mathscr{H}_{n}(z)}\psi_2^\pm(z),~~j=1,r.
\end{align}
Thus, we have
\begin{align}
  d \ln (\psi_1^\pm(z,x,y,t_{r+1}))=&~\left(\frac{\mp q \sqrt{\mathscr{R}_{2n+2}(z)}+\frac{1}{2}\mathscr{F}_{n,x}(z)}{\mathscr{F}_{n}(z)}\right)dx \nonumber\\
   &~+
  \left(\frac{\mp \widehat{F}_1(z) \sqrt{\mathscr{R}_{2n+2}(z)}+\frac{1}{2}\mathscr{F}_{n,y}(z)}{\mathscr{F}_{n}(z)}\right) dy\nonumber\\
  &~+
   \left(\frac{\mp \widehat{F}_2(z) \sqrt{\mathscr{R}_{2n+2}(z)}+\frac{1}{2}\mathscr{F}_{n,t_{r+1}}(z)}{\mathscr{F}_{n}(z)}\right) dt_{r+1}
\end{align}
and
\begin{align}
  d \ln (\psi_2^\pm(z,x,y,t_{r+1}))=&~\left(\frac{\pm p \sqrt{\mathscr{R}_{2n+2}(z)}+\frac{1}{2}\mathscr{H}_{n,x}(z)}{\mathscr{H}_{n}(z)}\right)dx \nonumber\\
   &~+
  \left(\frac{\mp \widehat{H}_1(z) \sqrt{\mathscr{R}_{2n+2}(z)}+\frac{1}{2}\mathscr{H}_{n,y}(z)}{\mathscr{H}_{n}(z)}\right) dy\nonumber\\
  &~+\left(
   \frac{\mp \widehat{H}_2(z) \sqrt{\mathscr{R}_{2n+2}(z)}+\frac{1}{2}\mathscr{H}_{n,t_{r+1}}(z)}{\mathscr{H}_{n}(z)}\right) dt_{r+1}.
\end{align}
According to relations (\ref{4.6kp})-(\ref{4.8kp}), it follows that
the integrals
\begin{equation}\label{4.18kp}
 \int_{(x_0,y_0,t_{r+1,0})}^{(x,y,t_{r+1})}d \ln \big(\frac{\psi_j^\pm(z,x,y,t_{r+1})}{\sqrt{\mathscr{F}_n(z,x,y,t_{r+1})}}\big),~~j=1,2
\end{equation}
is independent of the path. Therefore taking into account the normalization condition $\psi_1(P,x_0,y_0,t_{r+1,0})=1$, and
choosing
a special path
$$(x_0,y_0,t_{r+1,0})\rightarrow (x_0,y_0,t_{r+1})\rightarrow (x_0,y,t_{r+1}) \rightarrow (x,y,t_{r+1})$$
in (\ref{4.18kp}),
we finally obtain (\ref{4.10kp}) and (\ref{4.11kp}).  \qed\vspace{0.3cm}

Basic properties of Baker-Akhiezer
functions $\psi_j(P),j=1,2$ are summarized in the following result.

\newtheorem{lem4.4}[lem4.1]{Lemma}
\begin{lem4.4}
Suppose $q,p\in C^{\infty}(\mathbb{R}^{2+1})$ and $P\in X\backslash\{P_{\infty\pm}\}$.
Then the Baker-Akhiezer function derived in (\ref{4.10kp}), (\ref{4.11kp}) satisfy
\begin{align}
  &\psi_1(P,x,y,t_{r+1})\psi_1(P^*,x,y,t_{r+1})=\frac{\mathscr{F}_n(z,x,y,t_{r+1})}{\mathscr{F}_n(z,x_0,y_0,t_{r+1,0})}, \label{4.19kp}\\
  &\psi_2(P,x,y,t_{r+1})\psi_2(P^*,x,y,t_{r+1})=\frac{\mathscr{H}_n(z,x,y,t_{r+1})}{\mathscr{F}_n(z,x_0,y_0,t_{r+1,0})}, \label{4.20kp}\\
  & \psi_1(P,x,y,t_{r+1})\psi_2(P^*,x,y,t_{r+1})=\frac{\mathscr{G}_{n+1}(z,x,y,t)+\sqrt{\mathscr{R}_{2n+2}(z)}}{\mathscr{F}_{n}(z,x_0,y_0,t_{r+1,0})},\label{4.21kp}\\
  &\psi_1(P,x,y,t_{r+1})\psi_2(P^*,x,y,t_{r+1})+\psi_1(P^*,x,y,t_{r+1})\psi_2(P,x,y,t_{r+1})\nonumber\\
  &~~~~~~~~~~~~~~~~~~~~~~~~~~~~~~~=\frac{2\mathscr{G}_{n+1}(z,x,y,t_{r+1})}{\mathscr{F}_{n}(z,x_0,y_0,t_{r+1,0})},\label{4.22kp}\\
  &\psi_1(P,x,y,t_{r+1})\psi_2(P^*,x,y,t_{r+1})-\psi_1(P^*,x,y,t_{r+1})\psi_2(P,x,y,t_{r+1})\nonumber\\
  &~~~~~~~~~~~~~~~~~~~~~~~~~~~~~~~=\frac{2\sqrt{\mathscr{R}_{2n+2}(z)}}{\mathscr{F}_{n}(z,x_0,y_0,t_{r+1,0})}.\label{4.23kp}
\end{align}
\proof It can be easily seen that (\ref{4.19kp}) and (\ref{4.20kp}) hold by (\ref{4.10kp}) and (\ref{4.11kp}), respectively. Then from (\ref{4.19kp}), (\ref{4.20kp}) and the fact that $$\left(\frac{1}{2}(\psi_1(P)\psi_2(P^*)+\psi_1(P^*)\psi_2(P)), \psi_1(P)\psi_1(P^*), \psi_2(P)\psi_2(P^*)\right)$$
 and
 $$\left(\frac{\mathscr{G}_{n+1}(z,x,y,t_{r+1})}{\mathscr{F}_n(z,x_0,y_0,t_{r+1,0})}, \frac{\mathscr{F}_{}(z,x,y,t_{r+1})}{\mathscr{F}_n(z,x_0,y_0,t_{r+1,0})}, \frac{\mathscr{H}_{n}(z,x,y,t_{r+1})}{\mathscr{F}_n(z,x_0,y_0,t_{r+1,0})}\right)$$  are both solutions of
 linear system (\ref{2.16k})-(\ref{2.24k}) satisfying the same initial condition, one gets (\ref{4.22kp}). Using (\ref{4.19kp}), (\ref{4.20kp}) and (\ref{4.22kp}), we have
\begin{align}
  &\psi_1(P,x,y,t_{r+1})\psi_2(P^*,x,y,t_{r+1})-\psi_1(P^*,x,y,t_{r+1})\psi_2(P,x,y,t_{r+1})\nonumber\\
 =& \sqrt{(\psi_1(P)\psi_2(P^*)+\psi_1(P^*)\psi_2(P))^2-4\psi_1(P)\psi_1(P^*)\psi_2(P)\psi_2(P^*)}\nonumber\\
 =&\frac{2\sqrt{\mathscr{R}_{2n+2}(z)}}{\mathscr{F}_{n}(z,x_0,y_0,t_{r+1,0})}.
\end{align}
Finally, (\ref{4.21kp}) is the direct result of (\ref{4.22kp}), (\ref{4.23kp}).
\qed

\end{lem4.4}

 Next we consider the analytic property and asymptotic behavior of $\psi_1(P)\psi_2(P^*)$, and
 $\psi_j(P), j=1,2.$

\newtheorem{lem4.5a}[lem4.1]{Lemma}
\begin{lem4.5a}
The function $\psi_1(P)\psi_2(P^*)$ is a meromorphic function on $X$ with divisor
\begin{equation}\label{4.25dp}
(\psi_1(P)\psi_2(P^*))=\mathcal{D}_{\underline{\hat{\nu}}^*(x,y,t)\underline{\hat{\mu}}(x,y,t)P_{\infty+}}-\mathcal{D}_{\underline{\hat{\mu}}^*(x_0,y_0,t_0)\underline{\hat{\mu}}(x_0,y_0,t_0)P_{\infty-}},
\end{equation}
where we abbreviate $\underline{\hat{\mu}}=(\hat{\mu}_1,\ldots,\hat{\mu}_n), \underline{\hat{\nu}}=(\hat{\nu}_1,\ldots,\hat{\nu}_n).$
Moreover, using the local coordinate $\zeta=z^{-1}$ near $P_{\infty\pm},$ we have
\begin{align}\label{4.26dp}
   \psi_1(P,x,y,t)\psi_2(P^*,x,y,t)\overset{\zeta\rightarrow 0}{=}
   \begin{cases}
   \frac{1}{q(x_0,y_0,t_{r+1,0})}\zeta^{-1}+O(1), & \textrm{as}~~P\rightarrow P_{\infty+}, \cr
   - \frac{ p(x,y,t_{r+1})q(x,y,t_{r+1})}{ q(x_0,y_0,t_{r+1,0})}\zeta+O(\zeta^2), & \textrm{as}~~P\rightarrow P_{\infty-}. \cr
   \end{cases}
\end{align}
\end{lem4.5a}
\proof
Noticing  (\ref{3.3zp}), (\ref{2.22}), (\ref{4.21kp}) and
\begin{align*}
   \psi_1(P,x,y,t_{r+1})\psi_2(P^*,x,y,t_{r+1})=
   &~\frac{\mathscr{G}_{n+1}(z,x,y,t_{r+1})+\sqrt{\mathscr{R}_{2n+2}(z)}}{\mathscr{F}_{n}(z,x_0,y_0,t_{r+1,0})}\nonumber\\
   =&~\frac{\mathscr{G}_{n+1}(z,x,y,t_{r+1})+\sqrt{\mathscr{R}_{2n+2}(z)}}{\mathscr{F}_{n}(z,x,y,t_{r+1})}\frac{\mathscr{F}_{n}(z,x,y,t_{r+1})}{\mathscr{F}_{n}(z,x_0,y_0,t_{r+1,0})}\nonumber\\
   =&~\frac{\mathscr{H}_{n}(z,x,y,t_{r+1})}{\mathscr{G}_{n+1}(z,x,y,t_{r+1})-\sqrt{\mathscr{R}_{2n+2}(z)}}\frac{\mathscr{F}_{n}(z,x,y,t_{r+1})}{\mathscr{F}_{n}(z,x_0,y_0,t_{r+1,0})},\nonumber\\
\end{align*}
 one easily proves (\ref{4.25dp}) and (\ref{4.26dp}). \qed \vspace{0.3cm}

Now we turn to study the analytic structure of $\psi_j(P,x,y,t_{r+1})$ on $X\backslash\{P_{\infty\pm}\}.$

\newtheorem{thm4.5}[lem4.1]{Theorem}
\begin{thm4.5}
 Assume auxiliary spectrum points $\mu_j(x,y,t_{r+1})$, $\nu_j(x,y,t_{r+1}),$ \linebreak $j=1,\ldots,n,$ are mutually distinct and finite for all $(x,y,t_{r+1})\in\Omega,$
where $\Omega\in \mathbb{R}^3$ is an open interval.
Moreover, let $P\in X\backslash\{P_{\infty\pm}\}.$
Then
\begin{itemize}
\item[\textbf{I.}]
$\psi_1(P,x,y,t_{r+1})$ and $\psi_2(P,x,y,t_{r+1})$ are meromorphic on $P\in X\backslash\{P_{\infty\pm}\}.$ Their divisor of poles coincides with $\mathcal{D}_{\underline{\hat{\mu}}(x_0,y_0,t_{r+1,0})}.$

\item[\textbf{II.}] The divisor of zeros for $\psi_1(P,x,y,t_{r+1})$ and $\psi_2(P,x,y,t_{r+1})$
coincides with $\mathcal{D}_{\underline{\hat{\mu}}(x,y,t_{r+1})}$ and $\mathcal{D}_{\underline{\hat{\nu}}(x,y,t_{r+1})},$ respectively.

\item[\textbf{III.}] As $P\rightarrow P_{\infty\pm},$ the asymptotic behavior of $(\psi_1(P,x,y,t_{r+1}),\psi_2(P,x,y,t_{r+1}))^T$ is given by the equations,
    \begin{align}\label{4.27dp}
     \left(
       \begin{array}{c}
         \psi_1(P,x,y,t_{r+1}) \\
         \psi_2(P,x,y,t_{r+1}) \\
       \end{array}
     \right)=&~\left[\left(
                     \begin{array}{c}
                       0 \\
                       \frac{1}{q(x_0,y_0,t_{r+1,0})} \\
                     \end{array}
                   \right)\zeta^{-1}+\left(
                     \begin{array}{c}
                       \frac{q(x,y,t_{r+1})}{q(x_0,y_0,t_{r+1,0})} \\
                       O(1) \\
                     \end{array}
                   \right)+  O(\zeta)
     \right]\nonumber\\
     &\times\exp\Big(\frac{1}{2}(x-x_0)\zeta^{-1}-\frac{1}{2}(y-y_0)\zeta^{-2}\nonumber\\
     &-\frac{1}{2}(t_{r+1}-t_{r+1,0})\zeta^{-(r+1)}\Big),~~\textrm{at}~~P\rightarrow P_{\infty-},
    \end{align}
and
\begin{align}\label{4.28dp}
     \left(
       \begin{array}{c}
         \psi_1(P,x,y,t_{r+1}) \\
         \psi_2(P,x,y,t_{r+1}) \\
       \end{array}
     \right)=&~\left[\left(
                     \begin{array}{c}
                       1 \\
                       0 \\
                     \end{array}
                   \right)+
                   \left(
                     \begin{array}{c}
                       0 \\
                       -p(x,y,t_{r+1}) \\
                     \end{array}
                   \right)\zeta+O(\zeta^2)
                   \right]\nonumber\\
                   &\times\exp\Big(-\frac{1}{2}(x-x_0)\zeta^{-1}
     +\frac{1}{2}(y-y_0)\zeta^{-2}\nonumber\\
     &+\frac{1}{2}(t-t_{r+1,0})\zeta^{-(r+1)}\Big),~~\textrm{at}~~P\rightarrow P_{\infty+}.
    \end{align}
\end{itemize}
\end{thm4.5}

\proof First we study the function $\psi_1$. By (\ref{2.28l}), (\ref{2.29m}), (\ref{2.29n}) it follows
\begin{align}
 - \frac{1}{2}\frac{q(x^\prime,y,t_{r+1})y(P)}{\mathscr{F}_n(z,x^\prime,y,t_{r+1})}&\overset{P\rightarrow \hat{\mu}_j(x^\prime,y,t_{r+1})}{=} \partial_{x^\prime} \ln \sqrt{z-\mu_j(x^\prime,y,t_{r+1})},\\
  - \frac{y(P)}{2} \frac{\widehat{F}_1(z,x_0,y^\prime,t_{r+1})}{\mathscr{F}_n(z,x_0,y^\prime,t_{r+1})}&\overset{P\rightarrow \hat{\mu}_j(x_0,y^\prime,t_{r+1})}{=} \partial_{y^\prime} \ln \sqrt{z-\mu_j(x_0,y^\prime,t_{r+1})},\\
 - \frac{y(P)}{2} \frac{\widehat{F}_r(z,x_0,y_0,t^\prime)}{\mathscr{F}_n(z,x_0,y_0,t^\prime)}&\overset{P\rightarrow \hat{\mu}_j(x_0,y_0,t^\prime)}{=}~~ \partial_{t^\prime} \ln \sqrt{z-\mu_j(x_0,y_0,t^\prime)},
\end{align}
and hence one obtains
\begin{align}
 & \exp\Big(-\frac{1}{2}\int_{x_0}^x\frac{q(x^\prime,y,t_{r+1})y(P)}{\mathscr{F}_n(z,x^\prime,y,t_{r+1})}dx^\prime - \frac{y(P)}{2}\int_{y_0}^y\frac{\widehat{F}_1(z,x_0,y^\prime,t_{r+1})}{\mathscr{F}_n(z,x_0,y^\prime,t_{r+1})}dy^\prime\nonumber\\
 &-\frac{y(P)}{2}\int_{t_{r+1,0}}^{t_{r+1}}\frac{\widehat{F}_r(z,x_0,y_0,t^\prime)}{\mathscr{F}_n(z,x_0,y_0,t^\prime)}dt^\prime\Big)\nonumber\\
 =&\exp\Big(\int_{x_0}^x\partial_{x^\prime} \ln \sqrt{z-\mu_j(x^\prime,y,t_{r+1})}dx^\prime + \int_{y_0}^y
 \partial_{y^\prime} \ln \sqrt{z-\mu_j(x_0,y^\prime,t_{r+1})}dy^\prime\nonumber\\
 &+ \int_{t_{r+1,0}}^{t_{r+1}} \partial_{t^\prime} \ln \sqrt{z-\mu_j(x_0,y_0,t^\prime)}dt^\prime+O(1)
  \Big)\nonumber\\
 =&\begin{cases}
 \sqrt{z-\mu_j(x_0,y_0,t_{r+1,0})}^{-1},& P\rightarrow \hat{\mu}_j(x_0,y_0,t_{r+1,0}),\cr
 \sqrt{z-\mu_j(x,y,t_{r+1})},& P\rightarrow \hat{\mu}_j(x,y,t_{r+1})\neq \hat{\mu}_j(x_0,y_0,t_{r+1,0}),\cr
  O(1),& P\rightarrow \hat{\mu}_j(x,y,t_{r+1})= \hat{\mu}_j(x_0,y_0,t_{r+1,0}),\cr
  O(1),& P\rightarrow \textrm{other points}\neq  \hat{\mu}_j(x,y,t_{r+1}),~ \hat{\mu}_j(x_0,y_0,t_{r+1,0}).
 \end{cases}
 \end{align}
Then taking into account (\ref{4.10kp}), one proves {\bf I.} and {\bf II.} for $\psi_1$. Next we study the asymptotic behaviour of $\psi_1$ near $P_{\infty\pm}.$ Again using (\ref{4.10kp}) and local coordinate $\zeta={z}^{-1}$ near $P_{\infty\pm}$,
one infers
\begin{align}
\psi_1(P,x,y,t_{r+1})=&~\sqrt{\frac{\mathscr{F}_n(z,x,y,t_{r+1})}{\mathscr{F}_n(z,x_0,y_0,t_{r+1,0})}}
\exp\Big(-\frac{1}{2}\int_{x_0}^x\frac{q(x^\prime,y,t_{r+1})y(P)}{\mathscr{F}_n(z,x^\prime,y,t_{r+1})}dx^\prime \nonumber \\
&-\frac{y(P)}{2}\int_{y_0}^y\frac{\widehat{F}_1(z,x_0,y^\prime,t_{r+1})}{\mathscr{F}_n(z,x_0,y^\prime,t_{r+1})}dy^\prime
-\frac{y(P)}{2}\int_{t_{r+1,0}}^{t_{r+1}}\frac{\widehat{F}_r(z,x_0,y_0,t^\prime)}{\mathscr{F}_n(z,x_0,y_0,t^\prime)}dt^\prime\Big)\nonumber\\
=&\exp\Big(\int_{x_0}^x\left(\frac{-q(x^\prime,y,t_{r+1})y(P)}{2\mathscr{F}_n(z,x^\prime,y,t_{r+1})}
+\frac{\mathscr{F}_{n,x^\prime}(z,x^\prime,y,t_{r+1})}{2\mathscr{F}_n(z,x^\prime,y,t_{r+1})}\right)dx^\prime \nonumber \\
&+\int_{y_0}^y\left(-\frac{y(P)}{2}\frac{\widehat{F}_1(z,x_0,y^\prime,t_{r+1})}{\mathscr{F}_n(z,x_0,y^\prime,t_{r+1})}+ \frac{\mathscr{F}_{n,y^\prime}(z,x_0,y^\prime,t_{r+1})}{2\mathscr{F}_n(z,x_0,y^\prime,t_{r+1})}\right)dy^\prime \nonumber\\
&+\int_{t_{r+1,0}}^{t_{r+1}}\left(-\frac{y(P)}{2}\frac{\widehat{F}_r(z,x_0,y_0,t^\prime)}{\mathscr{F}_n(z,x_0,y_0,t^\prime)}
+\frac{\mathscr{F}_{n,t^\prime}(z,x_0,y_0,t^\prime)}{2\mathscr{F}_n(z,x_0,y_0,t^\prime)}\right)dt^\prime\Big)\nonumber\\
=&\begin{cases}
\exp\Big(\frac{1}{2}\int_{x_0}^x \left( \frac{-q(x^\prime,y,t_{r+1})}{\sum_{j=0}^\infty\hat{f}_j(x^\prime,y,t_{r+1})\zeta^{j+1}}
+\frac{q_{x^\prime}(x^\prime,y,t_{r+1})}{q(x^\prime,y,t_{r+1})}\right)dx^\prime\nonumber\\
~~~~+\frac{1}{2}\int_{y_0}^y \left( \frac{-\widehat{F}_1(x_0,y^\prime,t_{r+1})}{\sum_{j=0}^\infty\hat{f}_j(x_0,y^\prime,t_{r+1})\zeta^{j+1}}
+\frac{q_{y^\prime}(x_0,y^\prime,t_{r+1})}{q(x_0,y^\prime,t_{r+1})}\right)dy^\prime\nonumber\\
~~~~+\frac{1}{2}\int_{t_{r+1,0}}^{t_{r+1}} \left( \frac{-\widehat{F}_r(x_0,y_0,t^\prime)}{\sum_{j=0}^\infty\hat{f}_j(x_0,y_0,t^\prime)\zeta^{j+1}}+\frac{q_{t^\prime}(x_0,y_0,t^\prime)}{q(x_0,y_0,t^\prime)}\right)dt^\prime\Big),& \textrm{as}~~P\rightarrow P_{\infty-},
\cr
\exp\Big(\frac{1}{2}\int_{x_0}^x \left( \frac{q(x^\prime,y,t_{r+1})}{\sum_{j=0}^\infty\hat{f}_j(x^\prime,y,t_{r+1})\zeta^{j+1}}
+\frac{q_{x^\prime}(x^\prime,y,t_{r+1})}{q(x^\prime,y,t_{r+1})}\right)dx^\prime\nonumber\\
~~~~+\frac{1}{2}\int_{y_0}^y \left( \frac{\widehat{F}_1(x_0,y^\prime,t_{r+1})}{\sum_{j=0}^\infty\hat{f}_j(x_0,y^\prime,t_{r+1})\zeta^{j+1}}
+\frac{q_{y^\prime}(x_0,y^\prime,t_{r+1})}{q(x_0,y^\prime,t_{r+1})}\right)dy^\prime\nonumber\\
~~~~+\frac{1}{2}\int_{t_{r+1,0}}^{t_{r+1}} \left( \frac{\widehat{F}_r(x_0,y_0,t^\prime)}{\sum_{j=0}^\infty\hat{f}_j(x_0,y_0,t^\prime)\zeta^{j+1}}+\frac{q_{t^\prime}(x_0,y_0,t^\prime)}{q(x_0,y_0,t^\prime)}\right)dt^\prime\Big),&\textrm{as}~~P\rightarrow P_{\infty+},
\end{cases}\nonumber\\
=&\begin{cases}
\frac{q(x,y,t_{r+1})}{q(x_0,y_0,t_{r+1,0})}\exp\Big(\frac{1}{2}(x-x_0)\zeta^{-1}
 -\frac{1}{2}(y-y_0)\zeta^{-2}\nonumber\\~~~~~~~~~~~~~~~~~~~~~~~~~-\frac{1}{2}(t-t_{r+1,0})\zeta^{-(r+1)}+O(\zeta)\Big), ~~~~\textrm{as}~~P\rightarrow P_{\infty-},
\cr
\exp\Big(-\frac{1}{2}(x-x_0)\zeta^{-1}+\frac{1}{2}(y-y_0)\zeta^{-2}+\frac{1}{2}(t-t_{r+1,0})\zeta^{-(r+1)}+O(\zeta)\Big), \\ ~~~~~~~~~~~~~~~~~~~~~~~~~~~~~~~~~~~~~~~~~~~~~~~~~~~~~~~~~~~~~~~~~~\textrm{as}~~P\rightarrow P_{\infty+}.
\end{cases}\\\label{4.33dp}
\end{align}
Here we have used asymptotic spectral expansion
\begin{equation}
  \frac{\mathscr{F}_n(z,x,y,t_{r+1})}{y(P)}=\mp \sum_{j=0}^\infty\hat{f}_j(x,y,t_{r+1})\zeta^{j+1},~~\textrm{as}~~P\rightarrow P_{\infty\pm},
\end{equation}
which can be derived by induction as in \cite{Gesztesyhoden2003}. Applying Lemma 4.5
and (\ref{4.33dp}), we may derive related results for $\psi_2$. \qed

\section{Algebro-geometric solutions}

In this section, we shall first give a detailed description of
the function $\psi_1$
and then study
the theta function representation of Baker-Akhiezer functions $\psi_1(P), \psi_2(P)$ and algebro-geometric solutions of the KP hierarachy.

The function $\psi_1(P,x,y,t_{r+1})$  derived in last section plays very important roles.
Let us consider the case $r=2$ for example. By introducing
\begin{equation}\label{st5.1}
  \phi(P,x,y,t_{3})= \psi_1(P,x,-y,-t_{3})\exp(-\frac{1}{2}(x-x_0)z-\frac{1}{2}(y-y_0)z^2-\frac{1}{2}(t_3-t_{3,0})z^3),
\end{equation}
and using theorem 4.6, one infers the function $\phi$
possess the same properties {\bf I}, {\bf II} as $\psi_1$,
and the following expansions
\begin{equation}
  \phi(P,x,y,t_{3})\overset{\zeta\rightarrow 0}{=}
  \begin{cases}
  (1+O(\zeta))\exp(-(x-x_0)z-(y-y_0)z^2\\
  -(t_3-t_{3,0})z^3), & \textrm{as}~~P\rightarrow P_{\infty+},\cr
  O(1),&\textrm{as}~~ P\rightarrow P_{\infty-},
  \end{cases}
\end{equation}
where we use local coordinates $\zeta=z^{-1}$ near $P_{\infty\pm}.$
Thus, the function $\phi$ gives one explicit form of Baker-Akhiezer function for the KP equation
and one can use it
obtain the algebro-geometric solutions of the KP equation
following the way in \cite{Krichever}.

Another important fact is
the function $\psi_1(P,x,y,t_{3})$
describes a new (2+1) system, which is closely related with the KP equation.
Let
\begin{equation}
\Phi(P,x,y,t_{3})= \psi_1(P,x,y,t_{3})\exp(\frac{1}{2}(x-x_0)z-\frac{1}{2}(y-y_0)z^2-\frac{1}{2}(t_3-t_{3,0})z^3),
\end{equation}
and then we have the following results.

\newtheorem{thm5.1}{Theorem}[section]
\begin{thm5.1}
The function $\Phi(P,x,y,t_3)$ satisfies the following auxiliary linear problem:
\begin{equation}\label{5.1kp}
  L_2\Phi=0,~~L_3\Phi=0,
\end{equation}
where the operators $L_2$ and $L_3$ are given by the formulas
\begin{equation}\label{5.2zp}
\begin{split}
  L_2&= \partial_y+\partial^2_x+u_0\partial_x+u_1,\\
  L_3&= \partial_{t_3}+ \partial^3+v_0 \partial^2
  +v_1\partial_x+v_2,
\end{split}
\end{equation}
and
$u_0,u_1,v_0,v_1,v_2$ are coefficients independent of $P$
which are determined by the following conditions
\begin{align}
  (\partial_y-L_2)\Phi=&~O(\zeta)\exp\Big((x-x_0)\zeta^{-1} -(y-y_0)\zeta^{-2} - (t_3-t_{3,0})\zeta^{-3}\Big),\nonumber\\
  &~~~~~~~~~~~~~~~~~~~~~~~~~~~~~~~~\textrm{as}~~P\rightarrow P_{\infty-},~~\zeta=z^{-1},\label{5.5xyz}\\
  (\partial_{t_3}-L_3)\Phi=&~O(\zeta)\exp\Big((x-x_0)\zeta^{-1}-(y-y_0)\zeta^{-2}- (t_3-t_{3,0})\zeta^{-3}\Big),\nonumber\\
  &~~~~~~~~~~~~~~~~~~~~~~~~~~~~~~~~\textrm{as}~~P\rightarrow P_{\infty-},~~\zeta=z^{-1}.\label{5.6xyz}
 \end{align}

\end{thm5.1}
\proof
For convenience we denote by
\begin{equation}
 \Delta(x,y,t_3)=(x-x_0)\zeta^{-1} -(y-y_0)\zeta^{-2} - (t_3-t_{3,0})\zeta^{-3}
\end{equation}
and suppose
\begin{align}\label{5.5kp}
  \psi_1(P,x,y,t_3)\overset{\zeta\rightarrow 0}{=}&~\Big[\sum_{j=0}^\infty\Theta_j(x,y,t_3)\zeta^j\Big]e^{\Delta(x,y,t_3)},~~\textrm{as}~~P\rightarrow P_{\infty-}.
\end{align}
 Then it follows from (\ref{5.5kp}) that
\begin{align}\label{5.6zp}
  \psi_{1,x}(P,x,y,t_3)\overset{\zeta\rightarrow 0}{=}&~\Big[\Theta_0\zeta^{-1}+\sum_{j=0}^\infty(\Theta_{j,x}+\Theta_{j+1})\zeta^j\Big]e^{\Delta(x,y,t_3)},\\
   \psi_{1,xx}(P,x,y,t_3)\overset{\zeta\rightarrow 0}{=}&~\Big[\Theta_0\zeta^{-2}+( 2\Theta_{0,x}+\Theta_1)\zeta^{-1}+\sum_{j=0}^\infty(\Theta_{j,xx}+2\Theta_{j+1,x}\nonumber\\
   &+\Theta_{j+2})\zeta^j\Big]e^{\Delta(x,y,t_3)},\\
   \psi_{1,xxx}(P,x,y,t_3)\overset{\zeta\rightarrow 0}{=}&~
   ~\Big[\Theta_0\zeta^{-3}+ (3\Theta_{0,x}+\Theta_1)\zeta^{-2}+(3\Theta_{0,xx}+3\Theta_{1,x}+\Theta_2)\zeta^{-1}
   \nonumber\\
   &+\sum_{j=0}^\infty(\Theta_{j,xxx}+3\Theta_{j+1,xx}+3\Theta_{j+2,x}+\Theta_{3})\zeta^j\Big]e^{\Delta(x,y,t_3)}
  \end{align}
  and
 \begin{align}
  \psi_{1,y}(P,x,y,t_3)\overset{\zeta\rightarrow 0}{=}&~\Big[-\Theta_0 \zeta^{-2}-\Theta_1\zeta^{-1}+\sum_{j=0}^\infty(\Theta_{j,y}-\Theta_{j+2})\zeta^j\Big]e^{\Delta(x,y,t_3)}, \\
  \psi_{1,t_3}(P,x,y,t_3)\overset{\zeta\rightarrow 0}{=}&~\Big[-\Theta_0 \zeta^{-3}-\Theta_1\zeta^{-3}-\Theta_0\zeta^{-1}+\sum_{j=0}^\infty(\Theta_{j,t_3}-\Theta_{j+3})\zeta^j\Big]e^{\Delta(x,y,t_3)}.\label{5.10zp}
 \end{align}
Inserting (\ref{5.5kp})-(\ref{5.10zp}) into (\ref{5.1kp}), (\ref{5.2zp}), and taking into account (\ref{5.5xyz}), (\ref{5.6xyz}), one obtains
\begin{align}
  & 2\Theta_{0,x}+u_0\Theta_{0}=0,\label{5.15st}\\
  & \Theta_{0,y}+2 \Theta_{1,x}+ \Theta_{0,xx}+u_0 \Theta_{1}+u_1\Theta_{0}=0,\\
  &3 \Theta_{0,x}+v_0 \Theta_{0}=0,\\
  &3 \Theta_{1,x}+3 \Theta_{0,xx} +v_0 (\Theta_{1}+2 \Theta_{0,x})+v_1 \Theta_{0}=0,\\
  &3 \Theta_{2,x}+ 3\Theta_{1,xx}+ \Theta_{0,xxx} +v_0 (\Theta_{2}+2 \Theta_{1,x}+\Theta_{0,xx})\nonumber\\
  &
  +v_1(\Theta_{1}+\Theta_{0,x})+v_2 \Theta_{0}=0.\label{5.19st}
\end{align}
Here
$\Theta_{j}, j=0,1,2,\ldots,$ possess explicit representations which arise from (\ref{4.10kp})
or (\ref{5.15cp}). Thus solving the system (\ref{5.15st})-(\ref{5.19st}), we may derive
 $u_0,u_1,v_0,v_1,v_2$ which are expressed
 by $\Theta_{j}.$
 Since analytic properties of $L_2\psi_1$ and $L_3\psi_1$
are identical with $\psi_1$, except for possible different asymptotic behavior at $P_{\infty-}$, one
concludes
\begin{equation}
L_2\psi_1=0,~~L_3\psi_1=0
\end{equation}
hold by (\ref{5.5xyz}), (\ref{5.6xyz}).\qed\vspace{0.3cm}

\newtheorem{thm5.1new}[thm5.1]{Theorem}
\begin{thm5.1new}
Assume the condition of Theorem 5.1 holds. Then the compatibility condition for (\ref{5.1kp}), (\ref{5.2zp})
is equivalent to
\begin{align}
  &&2v_{0,x}-3u_{0,x}=0, \label{5.25xyz}\\
  &&v_{0,y}+v_{0,xx}+2 v_{1,x}-3u_{0,xx}-3u_{1,x}+u_0 v_{0,x}-2v_0 u_{0,x}=0,\\
 & &-u_{0,t_3}+v_{1,y}+v_{1,xx}+2v_{2,x}-u_{0,xxx}-3u_{1,xx}-2v_0 u_{1,x}
  +u_0v_{1,x}\nonumber\\
 & &-v_1 u_{0,x}-v_0 u_{0,xx}=0,
   \\
 & &-u_{1,t_3}+ v_{2,y}+v_{2,xx}+u_0 v_{2,x}-u_{1,xxx}-2v_0 u_{1,xx}-v_1u_{1,x}=0.\label{5.28xyz}
\end{align}

\end{thm5.1new}
\proof
Imposing on $\psi_1$ the requirement $\psi_{1,t_3y}=\psi_{1,yt_3}$,
one easily get (\ref{5.25xyz})-(\ref{5.28xyz}). \qed \vspace{0.3cm}

\newtheorem{wl}[thm5.1]{Remark}
\begin{wl}
The function $\psi_1(x,y,t_3)$ is connected with three equations, i.e.
the second and third flows of ANKS hierarchy, KP equation, and
the new system (\ref{5.25xyz})-(\ref{5.28xyz}). Thus, it is possible to
get the 'B\"{a}cklund transformation' of algebro-geometric solutions between
KP equation and the new 2+1 system. Similar statement is true for $r>2.$

\end{wl}

 Now we turn to study the theta function representations for $\psi_1(P), \psi_2(P)$
 and algebro-geometric solutions of the whole KP hierarchy.

First, choosing a convenient base point $Q_0 \in
     X \setminus \{P_{\infty\pm}\}$, the Abel maps
      $\underline{A}_{Q_0}(\cdot) $ and
      $\underline{\alpha}_{Q_0}(\cdot)$ are defined by
         \begin{eqnarray*}
           \underline{A}_{Q_0}:X \rightarrow
           J(X)&=&\mathbb{C}^{n}/L_{n},
         \end{eqnarray*}
         \begin{eqnarray*}
          P \mapsto \underline{A}_{Q_0} (P)&=& (A_{Q_0,1}(P),\ldots,
           A_{Q_0,n} (P)) \\
           &=&\left(\int_{Q_0}^P\omega_1,\ldots,\int_{Q_0}^P\omega_{n}\right)
           (\mathrm{mod}~L_{n}),
         \end{eqnarray*}
     and
         \begin{eqnarray*}
          && \underline{\alpha}_{Q_0}:
          \mathrm{Div}(X) \rightarrow
          J(X),\\
          &&~~~~~\qquad \mathcal{D} \mapsto \underline{\alpha}_{Q_0}
          (\mathcal{D})= \sum_{P\in \mathcal{K}_{n}}
           \mathcal{D}(P)\underline{A}_{Q_0} (P),
         \end{eqnarray*}
    where $L_{n}=\{\underline{z}\in \mathbb{C}^{n}|
           ~\underline{z}=\underline{N}+\Gamma\underline{M},
           ~\underline{N},~\underline{M}\in \mathbb{Z}^{n}\}$, and
  $\Gamma$, $\underline{\Xi}_{Q_0}$ are the Riemann matrix and the vector of Riemann
     constants, respectively. Moreover,
      we choose a homology basis $\{a_{j},b_{j}\}_{j=1}^{n}$ on $X$ in such a way that the intersection matrix of the cycles satisfies
\begin{equation}
a_{j}\circ b_{k}=\delta_{j,k}, \quad a_{j}\circ a_{k}=0,\quad b_{j}\circ b_{k}=0,\quad j,~k=1,\ldots,n.
\end{equation}
     For brevity, define the function
      $\underline{z}: X \times
      \sigma^{n} X \rightarrow \mathbb{C}^{n}$ by\footnote{$\sigma^{n} X$=
      $\underbrace{ X\times\ldots\times X}_{n}.$}
     \begin{eqnarray}\label{4.4}
           \underline{z}(P,\underline{Q})&=~\underline{\Xi}_{Q_0}
           -\underline{A}_{Q_0}(P)+\underline{\alpha}_{Q_0}
             (\mathcal{D}_{\underline{Q}}), \nonumber \\
           P\in \mathcal{K}_{n},\,~
           \underline{Q}&=~(Q_1,\ldots,Q_{n})\in
           \sigma^{n}\mathcal{K}_{n},
         \end{eqnarray}
     here $\underline{z}(\cdot,\underline{Q}) $ is
     independent of the choice of base point $Q_0$.
     The Riemann theta
     function $\theta(\underline{z})$ associated with $X$ and the homology is
      defined by
     $$\theta(\underline{z})=\sum_{\underline{n}\in\mathbb{Z}}\exp\left(2\pi i<\underline{n},\underline{z}>+\pi i<\underline{n},\underline{n}\Gamma>\right),\quad \underline{z}\in\mathbb{C}^{n},$$
     where $<\underline{B},\underline{C}>=\overline{\underline{B}}
     \cdot\underline{C}^t=\sum_{j=1}^{N}\overline{B}_jC_j$
     denotes the scalar product in $\mathbb{C}^{n-1}$.

Let $\omega_{P_{\infty\pm},q}^{(2)}$ be the normalized differentials
of the second kind with a unique pole at $P_{\infty\pm}$, respectively,
and principal parts
 \begin{equation}\label{5.13}
 \omega_{P_{\infty\pm},q}^{(2)}\underset{\zeta\rightarrow0}{=}\left(\zeta^{-2-q}+O(1)
\right)d\zeta,~~P\rightarrow P_{\infty\pm},~~\zeta=z^{-1},~~q\in\mathbb{N}_{0}
\end{equation}
with vanishing $a$-periods,
\begin{equation*}
\int_{a_{j}}\omega_{P_{\infty\pm},q}^{(2)}=0,~~ j=1,\ldots,n,
\end{equation*}
and
$\omega_{P_{\infty+}P_{\infty-}}$ be normalized differential of third kind
satisfying
\begin{equation*}
\begin{split}
   \int_{Q_0}^P\omega_{P_{\infty-}P_{\infty+}}=\zeta^{-1}+c_{\infty+}+O(\zeta),~~ P\rightarrow P_{\infty+}, \zeta=z^{-1}, c_{\infty+}\in\mathbb{C},
 \end{split}
 \end{equation*}
 and
 \begin{equation*}
 \begin{split}
 \int_{a_{j}}\omega_{P_{\infty+}P_{\infty-}}=0~~ j=1,\ldots,n.
  \end{split}
\end{equation*}
Moreover, we introduce the notation
 \begin{equation}\label{5.14}
\Omega_{r}^{(2)}=
 \frac{1}{2}(r+1)(\omega_{P_{\infty+},r}^{(2)}-\omega_{P_{\infty-},r}^{(2)}),~~r\in\mathbb{N}_0.
\end{equation}
and
\begin{equation}
 \omega_{r}^{(2)} =\lim_{P\rightarrow P_{\infty+}}\int_{Q_0}^P\Omega_{r}^{(2)}.
\end{equation}

Next we turn to study the theta function representation for $\psi_1(P)\psi_2(P^*)$.

\newtheorem{thm5.2}[thm5.1]{Lemma}
\begin{thm5.2}
Assume spectral curve defined in (\ref{2.22}) is nonsingular and $q,p$ satisfy (\ref{st2.51}).
Moreover, let $P\in X\backslash\{P_{\infty\pm}\}$ and suppose the divisors $\mathcal{D}_{\underline{\hat{\nu}}(x,y,t_{r+1})}$, $\mathcal{D}_{\underline{\hat{\mu}}(x,y,t_{r+1})}$ are nonspecial. Then  \\
\begin{align}\label{5.16a}
&\psi_1(P,x,y,t_{r+1})\psi_2(P^*,x,y,t_{r+1})\nonumber\\
=&\frac{1}{q(x_0,y_0,t_{r+1,0})}
\frac{\theta(\underline{z}(P_{\infty-}, \underline{\hat{\mu}}^*(x_0,y_0,t_{r+1,0})))\theta(\underline{z}(P_{\infty-}, \underline{\hat{\mu}}(x_0,y_0,t_{r+1,0})))}{\theta(\underline{z}(P_{\infty-}, \underline{\hat{\nu}}^*(x,y,t_{r+1})))\theta(\underline{z}(P_{\infty-}, \underline{\hat{\mu}}(x,y,t_{r+1})))}\nonumber\\
&\times \frac{\theta(\underline{z}(P, \underline{\hat{\nu}}^*(x,y,t_{r+1})))\theta(\underline{z}(P, \underline{\hat{\mu}}(x,y,t_{r+1})))}{\theta(\underline{z}(P, \underline{\hat{\mu}}^*(x_0,y_0,t_{r+1,0})))\theta(\underline{z}(P, \underline{\hat{\mu}}(x_0,y_0,t_{r+1,0})))}\nonumber
\\
&\times\exp\Big(\int_{Q_0}^P\omega_{P_{\infty-}P_{\infty+}}-c_{\infty+}\Big)\\
=&\frac{1}{q(x_0,y_0,t_{r+1,0})}\frac{\theta(\underline{z}(P_{\infty+}, \underline{\hat{\mu}}(x_0,y_0,t_{r+1,0})))\theta(\underline{z}(P_{\infty-}, \underline{\hat{\mu}}(x_0,y_0,t_{r+1,0})))}{\theta(\underline{z}(P_{\infty+}, \underline{\hat{\nu}}(x,y,t_{r+1})))\theta(\underline{z}(P_{\infty-}, \underline{\hat{\mu}}(x,y,t_{r+1})))}\nonumber\\
&\times \frac{\theta(\underline{z}(P^*, \underline{\hat{\nu}}(x,y,t_{r+1})))\theta(\underline{z}(P, \underline{\hat{\mu}}(x,y,t_{r+1})))}{\theta(\underline{z}(P^*, \underline{\hat{\mu}}(x_0,y_0,t_{r+1,0})))\theta(\underline{z}(P, \underline{\hat{\mu}}(x_0,y_0,t_{r+1,0})))}\nonumber
\\
&\times\exp\Big(\int_{Q_0}^P\omega_{P_{\infty-}P_{\infty+}}-c_{\infty+}\Big).
\end{align}

\end{thm5.2}

\proof 
We define
\begin{align}
 \circledR^1(P,x,t_{r+1})=&\frac{1}{q(x_0,y_0,t_{r+1,0})}
\frac{\theta(\underline{z}(P_{\infty-}, \underline{\hat{\mu}}^*(x_0,y_0,t_{r+1,0})))\theta(\underline{z}(P_{\infty-}, \underline{\hat{\mu}}(x_0,y_0,t_{r+1,0})))}{\theta(\underline{z}(P_{\infty-}, \underline{\hat{\nu}}^*(x,y,t_{r+1})))\theta(\underline{z}(P_{\infty-}, \underline{\hat{\mu}}(x,y,t_{r+1})))}\nonumber\\
&\times \frac{\theta(\underline{z}(P, \underline{\hat{\nu}}^*(x,y,t_{r+1})))\theta(\underline{z}(P, \underline{\hat{\mu}}(x,y,t_{r+1})))}{\theta(\underline{z}(P, \underline{\hat{\mu}}^*(x_0,y_0,t_{r+1,0})))\theta(\underline{z}(P, \underline{\hat{\mu}}(x_0,y_0,t_{r+1,0})))}\nonumber
\\
&\times\exp\Big(\int_{Q_0}^P\omega_{P_{\infty-}P_{\infty+}}-c_{\infty+}\Big)\\
 \circledR^2(P,x,t_{r+1})=&~\frac{1}{q(x_0,y_0,t_{r+1,0})}\frac{\theta(\underline{z}(P_{\infty+}, \underline{\hat{\mu}}(x_0,y_0,t_{r+1,0})))\theta(\underline{z}(P_{\infty-}, \underline{\hat{\mu}}(x_0,y_0,t_{r+1,0})))}{\theta(\underline{z}(P_{\infty+}, \underline{\hat{\nu}}(x,y,t_{r+1})))\theta(\underline{z}(P_{\infty-}, \underline{\hat{\mu}}(x,y,t_{r+1})))}\nonumber\\
&\times \frac{\theta(\underline{z}(P^*, \underline{\hat{\nu}}(x,y,t_{r+1})))\theta(\underline{z}(P, \underline{\hat{\mu}}(x,y,t_{r+1})))}{\theta(\underline{z}(P^*, \underline{\hat{\mu}}(x_0,y_0,t_{r+1,0})))\theta(\underline{z}(P, \underline{\hat{\mu}}(x_0,y_0,t_{r+1,0})))}\nonumber
\\
&\times\exp\Big(\int_{Q_0}^P\omega_{P_{\infty-}P_{\infty+}}-c_{\infty+}\Big).
\end{align}
Then it is not difficult to know divisors of $\psi_1(P,x,y,t_{r+1})\psi_2(P^*,x,y,t_{r+1})$ coincides with that of $\circledR^j(P,x,t_{r+1}),j=1,2,$ on $X$. Therefore, the quantity $\frac{\psi_1(P,x,y,t_{r+1})\psi_2(P^*,x,y,t_{r+1})}{\circledR^j(P,x,y,t_{r+1})}$ is a constant which is independent on $P$ and we denote it by $C(x,y,t_{r+1}).$ Then taking $P\rightarrow P_{\infty+}$ and using Lemma 4.5,
one obtains
$$C(x,y,t_{r+1})=\lim_{P\rightarrow P_{\infty+}}\frac{\psi_1(P,x,y,t_{r+1})\psi_2(P^*,x,y,t_{r+1})}{\circledR^j(P,x,y,t_{r+1})}=1,~~j=1,2,$$
and hence equality $\psi_1(P,x,y,t_{r+1})\psi_2(P^*,x,y,t_{r+1})=\circledR^j(P,x,y,t_{r+1}), j=1,2,$ hold for any $P\in X.$
We complete the proof.\qed
\\

Given these preparations, one finally derives the following theta function representation
for Baker-Akhiezer functions $\psi_j(P,x,y,t_{r+1}),j=1,2,$ and solutions $q(x,y,t_{r+1}),p(x,y,t_{r+1}).$\\

\newtheorem{thm5.3}[thm5.1]{Theorem}
\begin{thm5.3}
Assume spectral curve defined in (\ref{2.22}) is nonsingular and $q,p$ satisfy (\ref{st2.51}).
Moreover, let $P\in X\backslash\{P_{\infty\pm}\}$ and suppose the divisors $\mathcal{D}_{\underline{\hat{\nu}}(x,y,t_{r+1})}$, $\mathcal{D}_{\underline{\hat{\mu}}(x,y,t_{r+1})}$ are nonspecial.
Then functions $\psi_1,\psi_2,q,p$ have
the following theta function representations
\begin{align}
  \psi_1(P,x,y,t_{r+1})=&~C_1(x,y,t_{r+1})\frac{\theta(\underline{z}(P, \underline{\hat{\mu}}(x,y,t_{r+1})))}{\theta(\underline{z}(P, \underline{\hat{\mu}}(x_0,y_0,t_{r,0})))}\exp\Big((x-x_0)\int_{Q_0}^P\Omega^{(2)}_0 \nonumber\\
  &-(y-y_0)\int_{Q_0}^P\Omega^{(2)}_1-(t-t_{r+1,0})\int_{Q_0}^P\Omega^{(2)}_r\Big),\label{5.15cp}\\
   \psi_2(P,x,y,t_{r+1})=&~C_2(x,y,t_{r+1})\frac{\theta(\underline{z}(P, \underline{\hat{\nu}}(x,y,t_{r+1})))}{\theta(\underline{z}(P, \underline{\hat{\mu}}(x_0,y_0,t_{r+1,0})))}\exp\Big((x-x_0)\int_{Q_0}^P\Omega^{(2)}_0 \nonumber\\
   &-(y-y_0)\int_{Q_0}^P\Omega^{(2)}_1
   -(t-t_{r+1,0})\int_{Q_0}^P\Omega^{(2)}_2+\int_{Q_0}^P\omega_{P_{\infty+} P_{\infty-}}\Big),\label{5.16cp}\\
  =&~\widetilde{C}_2(x,y,t_{r+1})\frac{\theta(\underline{z}(P^*, \underline{\hat{\nu}}^*(x,y,t_{r+1})))}{\theta(\underline{z}(P^*, \underline{\hat{\mu}}^*(x_0,y_0,t_{r+1,0})))}\exp\Big((x-x_0)\int_{Q_0}^P\Omega^{(2)}_0 \nonumber\\
   &+(y-y_0)\int_{Q_0}^P\Omega^{(2)}_1
   +(t-t_{r+1,0})\int_{Q_0}^P\Omega^{(2)}_r+\int_{Q_0}^P\omega_{P_{\infty+} P_{\infty-}}\Big),\label{5.16acp}
\end{align}
and
\begin{align}
   q(x,y,t_{r+1})
 =&~q(x_0,y_0,t_{r+1,0}) \frac{\theta(\underline{z}(P_{\infty+}, \underline{\hat{\mu}}(x_0,y_0,t_{r+1,0})))}{\theta(\underline{z}(P_{\infty+}, \underline{\hat{\mu}}(x,y,t_{r+1})))}\nonumber\\
 &\times
  \frac{\theta(\underline{z}(P_{\infty-}, \underline{\hat{\mu}}(x,y,t_{r+1})))}{\theta(\underline{z}(P_{\infty-}, \underline{\hat{\mu}}(x_0,y_0,t_{r+1,0})))}
  \times \exp\Big(-2(x-x_0) \omega^{(2)}_0 \nonumber\\
  &+2(y-y_0)\omega^{(2)}_1
   +2(t-t_{r+1,0})\omega^{(2)}_r\Big), \label{st5.42}\\
    p(x,y,t_{r+1})
   =&~\frac{-e^{-2c_{\infty-}}}{q(x_0,y_0,t_{r+1,0})} \frac{\theta(\underline{z}(P_{\infty-}, \underline{\hat{\mu}}(x_0,y_0,t_{r+1,0})))}{\theta(\underline{z}(P_{\infty-}, \underline{\hat{\nu}}(x,y,t_{r+1})))} \nonumber\\
   &\times
  \frac{\theta(\underline{z}(P_{\infty+}, \underline{\hat{\nu}}(x,y,t_{r+1})))}{\theta(\underline{z}(P_{\infty+}, \underline{\hat{\mu}}(x_0,y_0,t_{r+1,0})))}
  \times \exp\Big(2(x-x_0) \omega^{(2)}_0  \nonumber\\
  &-2(y-y_0)\omega^{(2)}_1-2(t-t_{r+1,0})\omega^{(2)}_r\Big),\label{st5.43}
  \end{align}
where
\begin{align}
  C_1(x,y,t_{r+1})=&~\frac{\theta(\underline{z}(P_{\infty+}, \underline{\hat{\mu}}(x_0,y_0,t_{r+1,0})))}{\theta(\underline{z}(P_{\infty+}, \underline{\hat{\mu}}(x,y,t_{r+1})))}\times
  \exp\Big(-(x-x_0) \omega^{(2)}_0 \nonumber\\
  &~+(y-y_0)\omega^{(2)}_1
   +(t-t_{r+1,0})\omega^{(2)}_r\Big), \\
    C_2(x,y,t_{r+1})
  =&~\frac{e^{-c_{\infty-}}}{q(x_0,y_0,t_{r+1,0})}\frac{\theta(\underline{z}(P_{\infty-}, \underline{\hat{\mu}}(x_0,y_0,t_{r+1,0})))}{\theta(\underline{z}(P_{\infty-}, \underline{\hat{\nu}}(x,y,t_{r+1})))} \times \exp\Big((x-x_0) \omega^{(2)}_0\nonumber \\
 &
  -(y-y_0)\omega^{(2)}_1
  -(t-t_{r+1,0})\omega^{(2)}_r\Big),\\
   \widetilde{C}_2(x,y,t_{r+1})
 =&~\frac{e^{-c_{\infty-}}}{q(x_0,y_0,t_{r+1,0})}\frac{\theta(\underline{z}(P_{\infty+}, \underline{\hat{\mu}}^*(x_0,y_0,t_{r+1,0})))}{\theta(\underline{z}(P_{\infty+}, \underline{\hat{\nu}}^*(x,y,t_{r+1})))}\times \exp\Big((x-x_0) \omega^{(2)}_0 \nonumber \\
 & ~
  -(y-y_0)\omega^{(2)}_1
  -(t_{r+1}-t_{r+1,0})\omega^{(2)}_r\Big),
\end{align}
and the initial values $q(x_0,y_0,t_{r+1,0})$ and $p(x_0,y_0,t_{r+1,0})$ are constrained by
\begin{align}\label{st5.47}
   &q(x_0,y_0,t_{r+1,0})p(x_0,y_0,t_{r+1,0}) \nonumber\\
=&-\frac{\theta(\underline{z}(P_{\infty-}, \underline{\hat{\mu}}(x_0,y_0,t_{r+1,0})))}{\theta(\underline{z}(P_{\infty+}, \underline{\hat{\mu}}(x_0,y_0,t_{r+1,0})))}\frac{\theta(\underline{z}(P_{\infty+}, \underline{\hat{\nu}}(x_0,y_0,t_{r+1,0})))}{\theta(\underline{z}(P_{\infty+}, \underline{\hat{\nu}}(x_0,y_0,t_{r+1,0})))}e^{-2c_{\infty-}}.
\end{align}
Moreover, the Abel map linearizes the auxiliary divisors
       $\mathcal{D}_{\hat{\underline{\mu}}(x,y,t_{r+1})},
       \mathcal{D}_{\hat{\underline{\nu}}(x,y,t_{r+1})}$
       in the sense that
         \begin{align}
         \underline{\alpha}_{Q_0}(\mathcal{D}_{\underline{\hat{\mu}}(x,y,t_{r+1})})
         =&~\underline{\alpha}_{Q_0}(\mathcal{D}_{\underline{\hat{\mu}}(x_0,y_0,t_{r+1,0})})
         -\underline{U}_0^{(2)}(x-x_0)+\underline{U}_{1}^{(2)}(y-y_{0})\nonumber\\
         &+ \underline{U}_{r}^{(2)}(t_{r+1}-t_{r+1,0}),\label{5.24m}\\
         \underline{\alpha}_{Q_0}(\mathcal{D}_{\underline{\hat{\nu}}(x,y,t_{r+1})})
         =&~\underline{\alpha}_{Q_0}(\mathcal{D}_{\underline{\hat{\mu}}(x_0,y_0,t_{r+1,0})})
         -\underline{U}_0^{(2)}(x-x_0)+\underline{U}_{1}^{(2)}(y-y_{0})\nonumber\\
         &+\underline{U}_{r}^{(2)}(t_{r+1}-t_{r+1,0})+A_{Q_0}(P_{\infty-})-A_{Q_0}(P_{\infty+}).\label{5.25m}
       \end{align}

\end{thm5.3}
\proof  First, we prove expression (\ref{5.15cp}). Denote the right hand of (\ref{5.15cp}) by $\tilde{\psi}_1$, and then one finds $\tilde{\psi}_1$ is meromorphic on $X\backslash\{P_{\infty\pm}\}$
with simple zeros at $\hat{\mu}_j(x,y,t_{r+1}),j=1,\ldots,n,$ and simple poles at $\hat{\mu}_j(x_0,y_0,t_{r+1,0}),j=1,\ldots,n,$ by Riemann vanishing theorem. A comparison of (\ref{4.27dp}), (\ref{4.28dp}), (\ref{5.15cp}) for $\tilde{\psi}_1$, taking into (\ref{5.13}), (\ref{5.14}) shows $\psi_1$ and $\tilde{\psi}_1$
have identical exponential behavior up to order $O(1)$ near $P_{\infty\pm}.$ Thus,
$\psi_1$ and $\tilde{\psi}_1$ share the same singularities and zeros and the Riemann-Roch-type
uniqueness result then proves that $\psi_1$ and $\tilde{\psi}_1$ coincide up to
normalization and we denote this normalization constant by $C_1(x,y,t_{r+1}).$ Hence $\psi_1(P,x,y,t_{r+1})$ has the form of (\ref{5.15cp}).
Comparing (\ref{4.28dp}) with (\ref{5.15cp}) and taking into account the asymptotic behavior near $P_{\infty+}$, we have
\begin{align}
  C_1(x,y,t_{r+1})=&~\frac{\theta(\underline{z}(P_{\infty+}, \underline{\hat{\mu}}(x_0,y_0,t_{r+1,0})))}{\theta(\underline{z}(P_{\infty+}, \underline{\hat{\mu}}(x,y,t_{r+1})))}
  \exp\Big(-(x-x_0) \omega^{(2)}_0 \nonumber\\
  &~+(y-y_0)\omega^{(2)}_1
   +(t-t_{r+1,0})\omega^{(2)}_2\Big).
\end{align}
A comparison of (\ref{4.21kp}), (\ref{5.15cp}), (\ref{5.16cp}) and (\ref{5.16acp}) then yields
\begin{align}\label{5.21cc}
 C_2(x,y,t_{r+1})=&~ \frac{e^{-c_{\infty-}}}{q(x_0,y_0,t_{r+1,0})C_1(x,y,t_{r+1})}\frac{\theta(\underline{z}(P_{\infty-}, \underline{\hat{\mu}}(x_0,y_0,t_{r+1,0})))}{\theta(\underline{z}(P_{\infty-}, \underline{\hat{\nu}}(x,y,t_{r+1})))}\nonumber\\
 & ~\times
  \frac{\theta(\underline{z}(P_{\infty+}, \underline{\hat{\mu}}(x_0,y_0,t_{r+1,0})))}{\theta(\underline{z}(P_{\infty+}, \underline{\hat{\mu}}(x,y,t_{r+1})))}\nonumber\\
  =&~\frac{e^{-c_{\infty-}}}{q(x_0,y_0,t_{r+1,0})}\frac{\theta(\underline{z}(P_{\infty+}, \underline{\hat{\mu}}(x_0,y_0,t_{r+1,0})))}{\theta(\underline{z}(P_{\infty+}, \underline{\hat{\nu}}(x,y,t_{r+1})))}\times \exp\Big((x-x_0) \omega^{(2)}_0 \nonumber\\
 &
  -(y-y_0)\omega^{(2)}_1
  -(t-t_{r+1,0})\omega^{(2)}_r\Big)
 \end{align}
and
\begin{align}\label{5.21}
 \widetilde{C}_2(x,y,t_{r+1})=&~ \frac{e^{-c_{\infty-}}}{q(x_0,y_0,t_{r+1,0})C_1(x,y,t_{r+1})}\frac{\theta(\underline{z}(P_{\infty+}, \underline{\hat{\mu}}^*(x_0,y_0,t_{r+1,0})))}{\theta(\underline{z}(P_{\infty+}, \underline{\hat{\nu}}^*(x,y,t_{r+1})))} \nonumber\\
 & ~\times
  \frac{\theta(\underline{z}(P_{\infty+}, \underline{\hat{\mu}}(x_0,y_0,t_{r+1,0})))}{\theta(\underline{z}(P_{\infty+}, \underline{\hat{\mu}}(x,y,t_{r+1})))}\nonumber\\
 =&~\frac{e^{-c_{\infty-}}}{q(x_0,y_0,t_{r+1,0})}\frac{\theta(\underline{z}(P_{\infty+}, \underline{\hat{\mu}}^*(x_0,y_0,t_{r+1,0})))}{\theta(\underline{z}(P_{\infty+}, \underline{\hat{\nu}}^*(x,y,t_{r+1})))}\times \exp\Big((x-x_0) \omega^{(2)}_0 \nonumber\\
 &
  -(y-y_0)\omega^{(2)}_1
  -(t_{r+1}-t_{r+1,0})\omega^{(2)}_r\Big).
 \end{align}
On the other hand, from the asymptotic behavior of both sides in  (\ref{4.19kp}) and (\ref{4.20kp}) for $P\rightarrow P_{\infty+}$, one infers
\begin{align}
   q(x,y,t_{r+1})=&~q(x_0,y_0,t_{r+1,0}) C_1(x,y,t_{r+1})^2\frac{\theta(\underline{z}(P_{\infty-}, \underline{\hat{\mu}}(x,y,t_{r+1})))}{\theta(\underline{z}(P_{\infty-}, \underline{\hat{\mu}}(x_0,y_0,t_{r+1,0})))}\nonumber\\
   &\times
  \frac{\theta(\underline{z}(P_{\infty+}, \underline{\hat{\mu}}(x,y,t_{r+1})))}{\theta(\underline{z}(P_{\infty+}, \underline{\hat{\mu}}(x_0,y_0,t_{r+1,0})))} \nonumber\\
 =&~q(x_0,y_0,t_{r+1,0}) \frac{\theta(\underline{z}(P_{\infty+}, \underline{\hat{\mu}}(x_0,y_0,t_{r+1,0})))}{\theta(\underline{z}(P_{\infty+}, \underline{\hat{\mu}}(x,y,t_{r+1})))} \nonumber\\
 &\times
  \frac{\theta(\underline{z}(P_{\infty-}, \underline{\hat{\mu}}(x,y,t_{r+1})))}{\theta(\underline{z}(P_{\infty-}, \underline{\hat{\mu}}(x_0,y_0,t_{r+1,0})))}\times\exp\Big(-2(x-x_0) \omega^{(2)}_0 \nonumber\\
  &+2(y-y_0)\omega^{(2)}_1
   +2(t-t_{r+1,0})\omega^{(2)}_r\Big), \label{5.22abc}
   \end{align}
   and
   \begin{align}
    p(x,y,t_{r+1})=&~-q(x_0,y_0,t_{r+1,0}) C_2(x,y,t_{r+1})^2\frac{\theta(\underline{z}(P_{\infty-}, \underline{\hat{\nu}}(x,y,t_{r+1})))}{\theta(\underline{z}(P_{\infty-}, \underline{\hat{\mu}}(x_0,y_0,t_{r+1,0})))}\nonumber\\
   & \times
  \frac{\theta(\underline{z}(P_{\infty+}, \underline{\hat{\nu}}(x,y,t_{r+1})))}{\theta(\underline{z}(P_{\infty+}, \underline{\hat{\mu}}(x_0,y_0,t_{r+1,0})))} \nonumber\\
   =&~\frac{-e^{-2c_{\infty-}}}{q(x_0,y_0,t_{r+1,0})} \frac{\theta(\underline{z}(P_{\infty-}, \underline{\hat{\mu}}(x_0,y_0,t_{r+1,0})))}{\theta(\underline{z}(P_{\infty-}, \underline{\hat{\nu}}(x,y,t_{r+1})))}\nonumber\\
   &\times
  \frac{\theta(\underline{z}(P_{\infty+}, \underline{\hat{\nu}}(x,y,t_{r+1})))}{\theta(\underline{z}(P_{\infty+}, \underline{\hat{\mu}}(x_0,y_0,t_{r+1,0})))}\times\exp\Big(2(x-x_0) \omega^{(2)}_0\nonumber\\
  &-2(y-y_0)\omega^{(2)}_1
   -2(t-t_{r+1,0})\omega^{(2)}_r\Big).\label{5.22abcd}
  \end{align}
  Moreover, (\ref{st5.47}) follows from (\ref{5.22abc}) and (\ref{5.22abcd})
  by taking $(x,y,t_{r+1})=(x_0,y_0,t_{r+1,0}).$
 Finally,
the linearization property of the
Abel map in (\ref{5.24m}) and (\ref{5.25m}) is a standard investigation of the differentials $$\Omega_i(x,y,t_{r+1})=
d \ln(\psi_i(\cdot,x,y,t_{r+1})),~i=1,2,$$
or standard Langrange interpolation procedure (see \cite{Gesztesyhoden2003},~\cite{Novikov}). \qed \vspace{0.3cm}

\newtheorem{fin}[thm5.1]{Theorem}
\begin{fin}
Assume spectral curve defined in (\ref{2.22}) is nonsingular and $q,p$ satisfy (\ref{st2.51}).
Moreover, let $P\in X\backslash\{P_{\infty\pm}\}$ and the divisors $\mathcal{D}_{\underline{\hat{\nu}}(x,y,t_{r+1})}$, $\mathcal{D}_{\underline{\hat{\mu}}(x,y,t_{r+1})}$ are nonspecial. The algbro-geometric solutions of $r$th KP equation ($r\geq 2$)
possess the following two kinds of theta function representations
       \begin{align}
        u(x,y,t_{r+1})=&~\partial_{x}^2\ln\left(\theta(\underline{z}(P_{\infty+}, \underline{\hat{\mu}}(x,-y,(-1)^r t_{r+1})))\right)-\frac{\lambda_0}{2},\label{solution1}
        \end{align}
        and
        \begin{align}
        u(x,y,t_{r+1})=&\frac{\theta(\underline{z}(P_{\infty+},\underline{\hat{\mu}}(x,-y,(-1)^rt_{r+1}))}{\theta(\underline{z}(P_{\infty-},\underline{\hat{\mu}}(x,-y,(-1)^rt_{r+1}))}
        \frac{\theta(\underline{z}(P_{\infty-},\underline{\hat{\nu}}(x,-y,(-1)^rt_{r+1}))}{\theta(\underline{z}(P_{\infty+},\underline{\hat{\nu}}(x,-y,(-1)^rt_{r+1}))} \nonumber\\
        &\times e^{-2c_{\infty-}}, \label{solution2}
       \end{align}
where $\lambda_0\in\mathbb{C}.$
\end{fin}

\proof Firstly let us compute (\ref{solution1}).
 Using (\ref{4.1k}), one infers
 \begin{align}\label{5.32b}
 \psi_{1,xx}=\Big(\frac{z^2}{4}+\frac{z}{2}\frac{q_x}{q}+qp\Big)\psi_1+\frac{q_x}{q}\psi_{1,x}.
\end{align}
Suppose $\psi_1$ has the following expansions near $P\rightarrow P_{\infty+}$
\begin{small}
\begin{align}\label{5.24as}
  \psi_1(P,x,y,t_{r+1})\overset{\zeta\rightarrow 0}{=}&~\Big(1+\sum_{j=1}^\infty\Upsilon_j(x,y,t_{r+1})\zeta^j\Big)\exp\Big(-\frac{1}{2}(x-x_0) (\zeta^{-1}\nonumber\\
  &+\sum_{j=0}^{\infty}\lambda_j\zeta^{j+1})
  +\frac{1}{2}(y-y_0) (\zeta^{-2}+O(1)) \nonumber\\
  &+\frac{1}{2}(t_{r+1}-t_{r+1,0}) (\zeta^{-(r+1)}+O(1))\Big), ~~\zeta=z^{-1},
   \end{align}
   \end{small}
where the constants $\lambda_j, j=1,2,\ldots,$
arise from Abel differentials of the second kind.
Then we have
\begin{small}
\begin{align}
 \psi_{1,x}(P,x,y,t_{r+1})\overset{\zeta\rightarrow 0}{=}&~\Big(\sum_{j=1}^\infty\Upsilon_{j,x}(x,y,t_{r+1})\zeta^j\Big)\exp\Big(-\frac{1}{2}(x-x_0) (\zeta^{-1}+\sum_{j=0}^{\infty}\lambda_j\zeta^{j+1}) \nonumber\\
  &~+\frac{1}{2}(y-y_0) (\zeta^{-2}+O(1))
   +\frac{1}{2}(t_{r+1}-t_{r+1,0}) (\zeta^{-(r+1)}+O(1))\Big)\nonumber\\
  &~+\Big(1+\sum_{j=1}^\infty\Upsilon_j(x,y,t_{r+1})\zeta^j\Big)\exp\Big(-\frac{1}{2}(x-x_0) (\zeta^{-1}+\sum_{j=0}^{\infty}\lambda_j\zeta^{j+1}) \nonumber\\
  &~+\frac{1}{2}(y-y_0) (\zeta^{-2}+O(1))
   +\frac{1}{2}(t_{r+1}-t_{r+1,0}) (\zeta^{-(r+1)}+O(1))\Big)
   \nonumber\\
  &\times (-\frac{1}{2})\times\Big(\zeta^{-1}+\sum_{j=0}^{\infty}\lambda_j\zeta^{j+1}\Big),~~~~~\textrm{as}~~P\rightarrow P_{\infty+}, \label{5.25as}\\
  \psi_{1,xx}(P,x,y,t_{r+1})\overset{\zeta\rightarrow 0}{=}&~\Big(\sum_{j=1}^\infty\Upsilon_{j,xx}(x,y,t_{r+1})\zeta^j\Big)\exp\Big(-\frac{1}{2}(x-x_0) (\zeta^{-1}+\sum_{j=0}^{\infty}\lambda_j\zeta^{j+1}) \nonumber\\
  &~+\frac{1}{2}(y-y_0) (\zeta^{-2}+O(1))
   +\frac{1}{2}(t_{r+1}-t_{r+1,0}) (\zeta^{-(r+1)}+O(1))\Big)\nonumber\\
  &~+2\Big(\sum_{j=1}^\infty\Upsilon_{j,x}(x,y,t_{r+1})\zeta^j\Big)\exp\Big(-\frac{1}{2}(x-x_0) (\zeta^{-1}+\sum_{j=0}^{\infty}\lambda_j\zeta^{j+1}) \nonumber\\
  &~+\frac{1}{2}(y-y_0) (\zeta^{-2}+O(1))
   +\frac{1}{2}(t_{r+1}-t_{r+1,0}) (\zeta^{-(r+1)}+O(1))\Big)
   \nonumber\\
  &\times (-\frac{1}{2})\times\Big(\zeta^{-1}+\sum_{j=0}^{\infty}\lambda_j\zeta^{j+1}\Big)+
  \Big(1+\sum_{j=1}^\infty\Upsilon_j(x,y,t_{r+1})\zeta^j\Big)\nonumber\\
  &\times\exp\Big(-\frac{1}{2}(x-x_0) (\zeta^{-1}+\sum_{j=0}^{\infty}\lambda_j\zeta^{j+1})
  +\frac{1}{2}(y-y_0) (\zeta^{-2}+O(1))\nonumber\\
   &+\frac{1}{2}(t_{r+1}-t_{r+1,0}) (\zeta^{-(r+1)}+O(1))\Big)\times (-\frac{1}{2})^2\times\Big(\zeta^{-1}+\sum_{j=0}^{\infty}\lambda_j\zeta^{j+1}\Big)^2,\nonumber\\
   &~~~~~~~~~~~~~~~~~~~~~~~~~~~~~~~~~~~~~~~~~~~~~~\textrm{as}~~P\rightarrow P_{\infty+}.\label{5.26as}
   \end{align}
   \end{small}
Inserting (\ref{5.24as})-(\ref{5.26as}) into (\ref{5.32b}), we get
\begin{equation}\label{5.28abc}
  pq=\frac{\lambda_0}{2}-\Upsilon_{1,x}.
\end{equation}
Next one has to determine $\Upsilon_{1,x}.$ Using (\ref{5.24m}), (\ref{5.25m}), one derives
\begin{small}
\begin{align}\label{5.28as}
\psi_1(P,x,y,t_{r+1})\overset{\zeta\rightarrow 0}{=}& \frac{\theta(\underline{z}(P_{\infty+}, \underline{\hat{\mu}}(x_0,y_0,t_{r+1,0})))}{\theta(\underline{z}(P_{\infty+}, \underline{\hat{\mu}}(x,y,t_{r+1})))}
\frac{\theta(\underline{z}(P, \underline{\hat{\mu}}(x,y,t)))}{\theta(\underline{z}(P, \underline{\hat{\mu}}(x_0,y_0,t_0))}\nonumber\\
&\times \exp\Big(-\frac{1}{2}(x-x_0) (\zeta^{-1}+\sum_{j=0}^{\infty}\lambda_j\zeta^{j+1})
    +\frac{1}{2}(y-y_0) (\zeta^{-2}+O(1)) \nonumber\\
&+\frac{1}{2}(t_{r+1}-t_{r+1,0}) (\zeta^{-(r+1)}+O(1))\Big) \nonumber\\
\overset{\zeta\rightarrow 0}{=}&\left(1-\frac{\partial_{\underline{U}_0^{(2)}}\theta(\underline{z}(P_{\infty+}, \underline{\hat{\mu}}(x,y,t_{r+1}))}{\theta(\underline{z}(P_{\infty+}, \underline{\hat{\mu}}(x,y,t_{r+1})))}\zeta+O(\zeta^2)\right)\nonumber\\
&\times\left(1-\frac{\partial_{\underline{U}_0^{(2)}}\theta(\underline{z}(P_{\infty+}, \underline{\hat{\mu}}(x_0,y_0,t_0))}{\theta(\underline{z}(P_{\infty+}, \underline{\hat{\mu}}(x_0,y_0,t_0)))}\zeta+O(\zeta^2)\right)^{-1} \nonumber\\
&\times \exp\Big(-\frac{1}{2}(x-x_0) (\zeta^{-1}+\sum_{j=0}^{\infty}\lambda_j\zeta^{j+1})
    +\frac{1}{2}(y-y_0) (\zeta^{-2}+O(1)) \nonumber\\
&+\frac{1}{2}(t_{r+1}-t_{r+1,0}) (\zeta^{-(r+1)}+O(1))\Big), \nonumber\\
\overset{\zeta\rightarrow 0}{=}&\left(1
+\left(\frac{\partial_{\underline{U}_0^{(2)}}\theta(\underline{z}(P_{\infty+}, \underline{\hat{\mu}}(x_0,y_0,t_{r+1,0}))}{\theta(\underline{z}(P_{\infty+}, \underline{\hat{\mu}}(x_0,y_0,t_{r+1,0})))}
-\frac{\partial_{\underline{U}_0^{(2)}}\theta(\underline{z}(P_{\infty+}, \underline{\hat{\mu}}(x,y,t_{r+1}))}{\theta(\underline{z}(P_{\infty+}, \underline{\hat{\mu}}(x,y,t_{r+1})))}\right)\zeta+O(\zeta^2)\right)\nonumber\\
&\times \exp\Big(-\frac{1}{2}(x-x_0) (\zeta^{-1}+\sum_{j=0}^{\infty}\lambda_j\zeta^{j+1})
    +\frac{1}{2}(y-y_0) (\zeta^{-2}+O(1)) \nonumber\\
&+\frac{1}{2}(t-t_{r+1,0}) (\zeta^{-(r+1)}+O(1))\Big),
\end{align}
\end{small}
\noindent where $\partial_{\underline{U}_0^{(2)}}$ denotes the direction derivative of $\theta$ function along
vector $\underline{U}_0^{(2)}\in\mathbb{C}^n$ at the point $\underline{z}(P_{\infty+},\hat{\underline{\mu}}(x_0,y_0,t_{r+1,0}))$. Thus, comparing (\ref{5.24as})
with (\ref{5.25as}), we have
\begin{equation}\label{5.30abc}
  \Upsilon_{1,x}=-\partial_{x}^2\ln\left(\theta(\underline{z}(P_{\infty+}, \underline{\hat{\mu}}(x,y,t_{r+1})))\right).
\end{equation}
Employing (\ref{5.22abc}), (\ref{5.22abcd}), (\ref{5.28abc}), (\ref{5.30abc}) and Theorem 2.1, one obtains (\ref{solution1}). Finally,
(\ref{solution2}) is the direct result of (\ref{st5.42}), (\ref{st5.43}) and Theorem 2.1. \qed \vspace{0.3cm}

\newtheorem{fin2}[thm5.1]{Remark}
\begin{fin2}

Here we give a few remarks on the special case $r=2.$\\
(i)
In case $r=2,$ expressions (\ref{solution1}) and (\ref{solution2})
give rise to the algebro-geometric solutions of KP equation and
 real algebro-geometric solution of KP equation can easily derived from (\ref{solution2})
 by studying reduction condition $q=\pm\bar{p}.$
Moreover, The transformation $x\rightarrow ix, y\rightarrow iy, t_{3}\rightarrow it_{3}$
transforms (\ref{solution1}) and (\ref{solution2}) to algebro-geometric solutions of unstable version of KP equation (KP1 equation). \\
(ii) The spectral curve of the KP equation (\ref{1.1kp}) is hyperelliptic which compactified by two different points $P_{\infty\pm}$
at infinity. Similarly we can discuss algebro-geometric
solution of the KP hierarchy on some
specific trigonal curves following this way.
Moreover, if (\ref{solution1}) and (\ref{solution2}) are independent of $y$,
then we can derive theta function representations for algebro-geometric solutions
of KdV equation. A similar remark
applies also to the generation of the solutions of the KP equation from the
solutions of the Boussinesq equation, given the assumption that $u=u(x, y)$ and
the dependence of $\psi_j(x, y, t_3, P), j=1,2,$ on $t_3$ is purely exponential.\\
(iii) It is not difficult to verify the expression
 \begin{align}
   u(x,y,t_3)=&~-2q(x,y,t_3)p(x,y,t_3) \nonumber\\
   =&~2\partial_{x}^2\ln\left(\theta(\underline{z}(P_{\infty+}, \underline{\hat{\mu}}(x,y, t_{3})))\right)-\lambda_0
 \end{align}
 gives rise to the algebro-geometric solutions of standard KP equation \cite{Krichever}
\begin{equation}\label{solution3}
  u_t=\frac{1}{4}u_{xxx}+\frac{3}{2}uu_x+\frac{3}{4}\partial^{-1}_xu_{yy}.
\end{equation}

 \end{fin2}

\section{Outlook}
An important feature of this construction is that the Riemann surface $X$ is not completely arbitrary.
In fact, it has been proved that $N$-component vector nonlinear Schr\"{o}dinger hierarchies (VNLS) are contained within the KP hierarchy and general algebro-geometric solutions of the KP hierarchy
can be approximated by solutions of $N$-component VNLS hierarchies, in the limit of
large $N$ \cite{miller}. Therefore, how to
derive the algebro-geometric solutions of $N$-component vector nonlinear Schr\"{o}dinger hierarchies
by extending the method of Gesztesy, \emph{et al.} remains to be a nontrival work, which gives
a complete answer to algebro-geometric solutions of the whole KP hierarchy.

\section*{Acknowledgements}
The work described in this paper was supported by grants from the National
Science Foundation of China (Project No. 10971031), and the Shanghai
Shuguang Tracking Project (Project No. 08GG01).

\end{document}